\documentclass[aps,pra,twocolumn]{revtex4-1}
\usepackage{amsmath,amsfonts,amssymb}
\usepackage{graphicx,float,calc}
\usepackage{color,ulem}
\usepackage[breaklinks,colorlinks,urlcolor=blue,citecolor=blue,linkcolor=blue]{hyperref}

\begin{document}

\author{Yu-Biao Wu$^{1}$}
\author{Guang-Can Guo$^{1}$}
\author{Zhen Zheng$^2$}
\thanks{zhenzhen.dr@outlook.com}
\author{Xu-Bo Zou$^{1}$}
\thanks{xbz@ustc.edu.cn}
\affiliation{$^1$Key Laboratory of Quantum Information, and CAS Center for Excellence in Quantum Information and Quantum Physics, University of Science and Technology of China, Hefei, Anhui 230026, People's Republic of China}
\affiliation{$^2$Department of Physics and HKU-UCAS Joint Institute for Theoretical and Computational Physics at Hong Kong, The University of Hong Kong, Pokfulam Road, Hong Kong, China}
\title{Effective Hamiltonian with tunable mixed pairing in driven optical lattices}

\begin{abstract}

Mixed pairing in ultracold Fermi gases can give rise to interesting many-body phases,
such as topological nontrivial superfluids that support Majorana zero modes (MZMs) with various spatial configurations.
Unfortunately, in ordinary lattice systems, the topological phase and the associated MZMs are suppressed by the dominant $s$-wave pairing.
Here we present a proposal for engineering effective Hamiltonians with tunable mixed on- and off-site pairing based on driven optical lattices.
The on- and off-site pairing can be changed independently by means of a periodical driving field rather than magnetic Feshbach resonances.
It paves the way for suppressing the dominant on-site interaction that frustrates the emergence of topological superfluids and for synthesizing MZMs localized in edges or corners.

\end{abstract}
\maketitle

\section{Introduction}
Topological superconductors have attracted intensive interest of condensed-matter as well as ultracold-atom physics in recent years.
Unlike the conventional superconductors, topological superconductors serve as promising candidates to observe Majorana zero modes (MZMs), with potential applications for fault-tolerant quantum computing \cite{quant-compu-rev}.
Based on the mathematical structure of Bogoliubov-de Gennes (BdG) Hamiltonians, a complete classification has been proposed for characterizing different kinds of topological superconductors as well as ways to engineer them in real experiments \cite{topo-cri-rev-1,topo-cri-rev-2}.
For example, topological superconductors with triplet pairing (e.g., the chiral $p$-wave pairing which breaks the time-reversal symmetry \cite{kitaev-p-wave}) can be engineered by
Rashba spin-orbit coupling (SOC) in presence of ordinary $s$-wave pairing \cite{soc-p-wave}.
On the other hand, unconventional singlet pairing (e.g., the $s_\pm$-wave or $d$-wave pairing)
can give rise to time-reversal-invariant (TRI) topological superconductors \cite{time-sc-Deng,time-sc-Zhang,time-sc-Keselman}, or ones characterized by higher-order topological invariants \cite{zhang-corner,wang-corner,zhu-corner-2018,liu-corner,vincent-corner,zhu-corner-2019,luo-corner-2019,yan-corner-2019}.
However, they are not easily accessible in conventional solid-state systems.

Compared with conventional solid-state systems,
ultracold atoms in optical lattices offer a remarkable platform for investigating quantum many-body problems \cite{quantum-simu-rev-nphys,quantum-simu-rev-sci}.
Typically, the optical lattices are constructed by interfering several laser beams, thus a fully controllable lattice geometry and tunable lattice depth is attainable.
Effective physical fields, for example, Zeeman fields and SOC, can be synthesized by lasers,
and their strengths are also tunable \cite{laser-hop-rev-1,laser-hop-rev-2}.
The control over many-body interactions can be achieved via Feshbach resonances,
and controlled by external magnetic or optical fields \cite{feshbach-rev-1,feshbach-rev-2}.
These technical advances have enabled the realization of superfluid neutral atomic Fermi gases \cite{fermi-gas-rev,ketterle2008proceedings}.
This motivates us to search a possible proposal for realizing tunable unconventional singlet pairing in Fermi gases.

Intuitively, unconventional singlet pairing can be introduced by two-body interactions with higher-partial-wave symmetries
(e.g., $d$-wave ones \cite{cold-atom-exp-d-wave}),
the realization of which near Feshbach resonances has, however, encountered great difficulties.
This is because the severe atomic loss prohibits the many-body equilibration in a reasonably long time scale.
An alternative scheme is based on engineering mixed on- and off-site interactions \cite{qi-d-wave,njp-d-wave,d-wave-NNN-1,d-wave-NNN-2}.
By introducing background bosonic molecules with macroscopic occupation in the ground state, the mixed pairing can be obtained by coupling two atoms to one molecule.
In these schemes, the on-site and off-site pairing arise by loading atoms or molecules into
a state-dependent optical lattice.
However, their strengths are simultaneously determined by the atom-molecule coupling,
and can not be independently controlled.

In order to generate independently tunable mixed pairing,
we propose a scheme based on Floquet engineering
\cite{driven-theo-1,driven-theo-2,driven-theo-3,floquet-rev,driven-Eckardt-2005,driven-Lignier-2007,lindner2011floquet,driven-Liu-2012,driven-Zheng-2014,driven-Struck-2014,driven-Zheng-2015,driven-Zheng-2016,driven-Meinert-2016,driven-Nocera-2017,driven-Xu-2017,driven-Grass-2018,driven-Zhou-2018,driven-Messer-2018,Zheng_2019} in this paper.
Floquet engineering has proven to be a versatile method for realizing a variety of 
unconventional effective Hamiltonians with tunable parameters,
for instance, correlated tunneling \cite{driven-Meinert-2016,driven-cor-hop-1},
spin-exchange interaction \cite{driven-Chen-2011,driven-Bukov-2016},
and artificial gauge fields \cite{driven-Zhang-2017,driven-Botao-2018}.
Here we report that, by introducing periodical driving external fields,
the strengths of mixed pairing can be controlled.
This makes it possible to individually tune the on- and off-site pairing strengths,
thus potentially synthesizing MZMs with various spatial configurations.

The paper is organized as follows.
In Sec.\ref{sec-model}, we describe the general model for tunable on- and off-site pairing based on the driving field.
Then in Sec.\ref{sec-app},
we present the applications of the tunable interaction in single-layer and bilayer systems,
showing the engineering of edge and corner MZMs.
In Sec.\ref{sec-exp}, the experimental realization of our proposal is discussed.
In Sec.\ref{sec-conclude}, we summarize the paper.

\section{Model Hamiltonian}\label{sec-model}

We consider ultracold Fermi gases loaded in a two-dimensional (2D) optical lattice.
The atomic interaction is controlled via Feshbach resonances.
It describes atomic Fermi gases in which two fermionic atoms (open channel)
are coupled to a bosonic molecular state (closed channel).
In our proposal, the fermions and bosonic states are confined in lattice potentials
$V_{F}({\bf r})=V_{F}[\sin^2(k_L x)+\sin^2(k_L y)]$ and $V_{B}({\bf r})=V_{B}[\sin^2(k_L x)+\sin^2(k_L y)]$,
with $k_L=\pi/a$ and $a$ being the lattice constant.
The interaction Hamiltonian can be determined by a two-channel model \cite{feshbach-rev-2},
which is formulated as
\begin{equation}
H_{\rm int}=g\int {\rm d}{\bf r}\, \psi_B^\dag({\bf r})\psi_\uparrow({\bf r}) \psi_\downarrow({\bf r}) +{\rm H.c.}
\end{equation}
Here $\psi_\sigma$ and $\psi_B$ are
operators for fermions of spin-$\sigma$ and bosonic states, respectively.
$g$ is the bare interaction strength in free space,
and H.c. stands for the Hermitian conjugation.

\begin{figure}[t]
\centering
\includegraphics[width=0.45\textwidth]{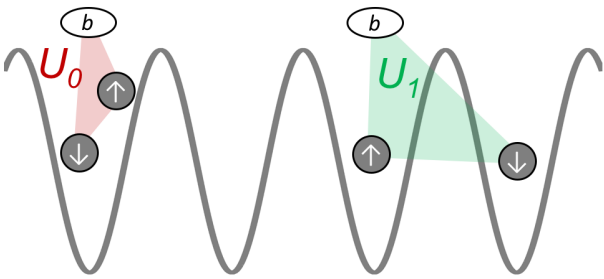}
\caption{Illustration of the interaction between the bosonic molecule state $b$ and spin-$\uparrow\downarrow$ fermions: 
the on-site one ($U_0$), and the off-site one ($U_1$).} 
\label{fig-model}
\end{figure}

We use the tight-binding approximation (TBA) to study the system.
The interaction Hamiltonian is expanded in terms of Wannier wave functions $W({\bf r})$ and $W_B({\bf r})$,
\begin{equation}
H_{\rm int} = \sum_{j,l}U_0b_{j}^\dag c_{j\uparrow} c_{j\downarrow}
+\frac{U_1}{2} (b_{j}^\dag+b_{j+{\bf e}_l}^\dag) c_{j\uparrow} c_{j+{\bf e}_l,\downarrow} +{\rm H.c.}
\label{eq-h-int}
\end{equation}
where $c_{\sigma}$ and $b$ are operators of fermions and bosonic states, respectively.
We write the site index as $j=(j_x,j_y)$ and ${\bf e}_{l=x,y}$ denotes the unit vectors of the primitive cell for fermions.
The interaction strengths are given by
\begin{equation}
\left\{\begin{split}
& U_0 = g\int {\rm d}{\bf r}\, W_B^*({\bf r}) W({\bf r})W({\bf r}) \\
& U_1 = 2g\int {\rm d}{\bf r}\, W_B^*({\bf r}) W({\bf r})W({\bf r}+a) 
\end{split}\right. \,. \label{eq-U-strength}
\end{equation}

\begin{figure}[t]
\centering
\includegraphics[width=0.5\textwidth]{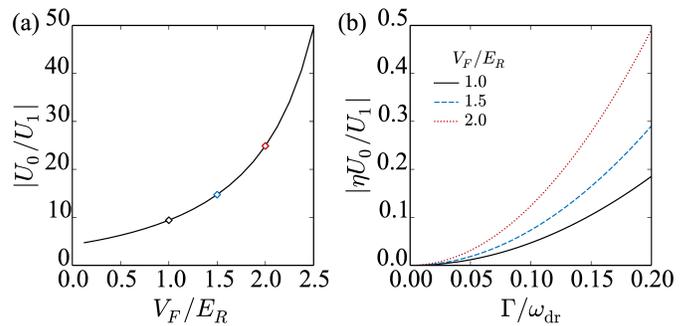}
\caption{(a) The ratio of bare interaction strengths $|U_0/U_1|$ as functions of the fermionic lattice trap depth $V_F$.
(b) The ratio of effective interaction strengths $|\eta U_0/U_1|$ as functions of $\Gamma/\omega_{\rm dr}$ 
for various $V_F$.
The bosonic lattice trap depth $V_B=2V_F$ \cite{lattice-feshbach-1,lattice-feshbach-2}.
Here $E_R=\hbar^2 k_L^2 /2m$ is the recoil energy of lattices.
Diamonds in (a) correspond to the lines with the same color in (b).} 
\label{fig-interaction}
\end{figure}

In Hamiltonian (\ref{eq-h-int}),
we have accounted for the off-site pairing, as shown in FIG. \ref{fig-model}.
Generally, $U_1$ is much smaller than $U_0$ (see FIG. \ref{fig-interaction}(a)).
Hence the on-site pairing is dominant, and the off-site one can be totally neglected
and does not bring in interesting physics.
One can design the lattice potential or apply magnetic Feshbach resonances to control
the profile of the on- and off-site interactions.
However, from Eq.(\ref{eq-U-strength}), we know that both on- and off-site terms are solely determined by the bare interaction $g$.
It reveals that the magnitude $|U_0/U_1|$ is independent of $g$.
However, as shown in previous works, single-particle terms such as the hopping magnitude
can be modified by Floquet engineering.
This inspires us to search for a possible routine to suppress the dominant on-site interaction and
design independently tunable mixed pairing.

We introduce a periodical driving term to the Hamiltonian (\ref{eq-h-int}),
\begin{equation}
H_{\rm int}(t) = H_{\rm int} + H_{\rm dr}(t) \,, \label{eq-h-total-start}
\end{equation}
where $H_{\rm dr}(t)=\sum_{j,\sigma} V_{\rm dr}(t) c_{j\sigma}^\dag c_{j\sigma}$.
$V_{\rm dr}(t)$ is a locally and periodical driving potential of the form
\begin{equation}
V_{\rm dr}(t) = \Gamma\cos(\omega_{\rm dr} t)\cos( k_{\rm dr}x+k_{\rm dr}y) + \nu_j
\,. \label{eq-driven-field}
\end{equation}
Here $\Gamma$ is the amplitude of the driving field.
$\nu_j$ is the spatially modulated energy offset
and can be engineered in a checkerboard structure $\nu_{j} = (-1)^{j_x+j_y}\omega_{\rm dr}$.
The parameters $\omega_{\rm dr}$ and $k_{\rm dr}$ are determined by the lasers that generate the driving field.
In TBA, we can obtain $k_{\rm dr}l=j_la$ ($l=x,y$) as long as we adjust $k_{\rm dr}$ to match the lattice vector $k_L$.
In order to get a time-dependent effective Hamiltonian,
we make the following rotation transformation,
\begin{equation}
\mathcal{U} = \exp [ i\int_{t_i}^{t} H_{\rm dr}(t') {\rm d}t' ]
\equiv e^{i\mathcal{\hat{A}}(t)} \label{eq-rotate}
\end{equation}
with $\mathcal{\hat{A}}(t)=\sum_{j,\sigma}\mathcal{A}(j,t)c_{j\sigma}^\dag c_{j\sigma}$ and $\mathcal{A}(j,t)=\frac{\Gamma}{\omega_{\rm dr}} (-1)^{j_x+j_y}\sin(\omega_{\rm dr} t)+\nu_{j} t$.
In the rotating frame, Hamiltonian (\ref{eq-h-total-start}) becomes
$H_{\rm int}(t)\rightarrow \mathcal{H}_{\rm int}(t)=\mathcal{U}H_{\rm int}(t)\mathcal{U}^\dag - i \mathcal{U} \partial_t \mathcal{U}^\dag$.
In this way, the time-dependent term $H_{\rm dr}(t)$ can be rotated off.

Due to the relation $\mathcal{A}(j+{\bf e}_{l},t)=-\mathcal{A}(j,t)$,
the $U_1$ term in Eq.(\ref{eq-h-int}) is unchanged under the transformation into the rotating frame of Eq.(\ref{eq-rotate}).
By contrast, $U_0$ is replaced by a time-dependent form:
$\tilde{U}_0(t) = U_0\sum_{n} \mathcal{J}_n(2\Gamma/\omega_{\rm dr})e^{i\tilde{\phi}(n,j,t)}$.
Here $\mathcal{J}_n(\cdot)$ stands for the Bessel function of the $n$-th order, and
$\tilde{\phi}(n,j,t) = n(-1)^{j_x+j_y}\omega_{\rm dr} t+2(-1)^{j_x+j_y}\omega_{\rm dr} t$.
We notice that the phase $\tilde{\phi}(n,j,t)=0$ only when $n=-2$, 
and the Bessel function obeys $\mathcal{J}_{-2}(\cdot)=\mathcal{J}_2(\cdot)$.
By neglecting rapidly oscillating terms,
the final form of the effective interaction Hamiltonian is expressed as
\begin{equation}
H_{\rm eff} = 
\sum_j  \eta U_0 b_{j}^\dag c_{j\uparrow} c_{j\downarrow}
+ \frac{U_1}{2} (b_{j}^\dag+b_{j+{\bf e}_l}^\dag) c_{j\uparrow} c_{j+{\bf e}_l,\downarrow}+ \text{H.c.} \label{eq-h-model-eff}
\end{equation} 
with $\eta=\mathcal{J}_2(2\Gamma/\omega_{\rm dr})$.

In Hamiltonian (\ref{eq-h-model-eff}),
the periodical driving potential $V_\mathrm{dr}(t)$
gives rise to a modified magnitude of the on-site interaction strength $U_0$,
while the off-site interaction strength $U_1$ remains unchanged.
As $\Gamma$ is a fully controllable parameter in real experiments,
it offers a feasible tool to change the ratio between the effective on-site ($\eta U_0$) and off-site interaction ($U_1$) strengths.
In FIG. \ref{fig-interaction}(b), we plot the ratio of effective interaction strengths $\eta U_0/U_1$ when changing $\Gamma/ \omega_\mathrm{dr}$.
In particular, when we prepare $\Gamma\ll \omega_\mathrm{dr}$,
it leads to $\eta \sim 0$.
Therefore, as shown in FIG. \ref{fig-interaction}, the off-site interaction strength can be dominant over the effective on-site one, 
even though the bare strength $U_0$ is much larger than $U_1$.

\section{Applications}\label{sec-app}

Next, we present two examples, 
by using the effective Hamiltonian (\ref{eq-h-model-eff}), for realizing specific MZMs with various spatial configurations.

\subsection{Edge MZM}
It is known that the TRI topological superfluid phase \cite{time-sc-Zhang}
has the following feature:
when the off-site pairing strength $U_1$ exceeds a critical threshold which depends on $U_0$,
the band gap closes and reopens, resulting in a topological phase transition 
from the trivial superfluid phase to a topological nontrivial one.
As the strength of $U_0$ and $U_1$ can be individually controlled,
the effective Hamiltonian (\ref{eq-h-model-eff}) is a promising candidate for the TRI topological superfluid phase.
This can be realized if we simultaneously generate a Rashba-type SOC,
\begin{equation}
H_{\rm soc} = i\alpha\sum_{j,\tau\tau'} (c_{j\tau}^\dag [s_x]_{\tau\tau'} c_{j+{\bf e}_y\tau'}
-c_{j\tau}^\dag [s_y]_{\tau\tau'} c_{j+{\bf e}_x\tau'}) + \text{H.c.}
\end{equation}
where $\alpha$ is the SOC strength,
$s_{x,y,z}$ are Pauli matrices in the spin space,
and $\tau=1$ and 2, respectively, stand for spin $\uparrow$ and $\downarrow$.
The total Hamiltonian in TBA is expressed as
\begin{equation}
H_{\rm 2D} = H_F + H_B + H_{\rm soc} + H_{\rm int} \,,
\end{equation}
where
\begin{align}
& H_F = -\sum_{\langle ij \rangle,\sigma} J c_{i\sigma}^\dag c_{j\sigma}
-\sum_{j,\sigma} \mu c_{j\sigma}^\dag c_{j\sigma} \,,\\
& H_B = -\sum_{\langle ij \rangle,\sigma} J_B b_{i}^\dag b_{j}
-\sum_{j,\sigma} \mu_B b_{j}^\dag b_{j} \,.
\end{align}
Here $H_F$ and $H_B$ describe the single-particle Hamiltonians of fermions and bosonic states originating from kinetic motion.
$\mu$ and $\mu_B=2\mu-\nu_0$ are the corresponding chemical potentials,
and $J$ and $J_B$ are the hopping magnitudes.
In $\mu_B$, $\nu_0$ is the bare detuning between the open and closed channels that is controllable in real experiments,
and $2\mu$ is imposed for the sake of number conservation.
By making the rotation transformation (see Eq. (\ref{eq-rotate})),
the forms of $H_F$ and $H_{\rm soc}$ remain unchanged except the hopping $J$ and SOC strength $\alpha$ are replaced by a modified magnitude: $J \rightarrow \eta J$ and $\alpha \rightarrow \eta \alpha$.
Therefore, the total effective Hamiltonian is given by
\begin{equation}
\mathcal{H}_{\rm 2D} = \eta H_F + H_B + H_{\rm eff} + \eta H_{\rm soc} \,. \label{eq-h-app-edge}
\end{equation}

\begin{figure}[t]
\centering
\includegraphics[width=0.5\textwidth]{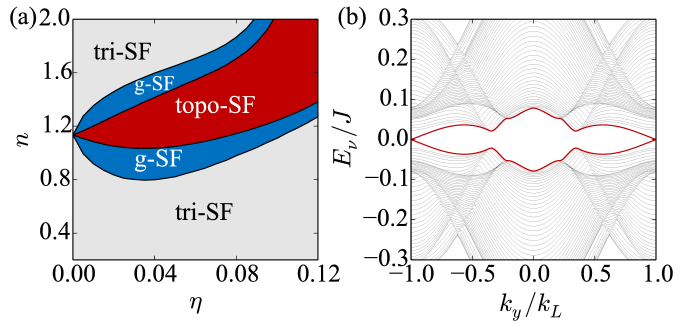}
\caption{(a) Phase diagram at zero temperature.
topo-SF, g-SF, and tri-SF stand for topological, gapless, and trivial superfluid phases, respectively.
We set $U_0=5.0J$, $U_1=0.32J$, $\nu_0=1.0J$, and $\alpha=0.5J$.
(b) BdG spectrum of the lattice system at $(\eta,n)=(0.06,1.2)$.
We use the open boundary condition in the $x$ direction with $L=100$
and periodical boundary condition in the $y$ direction.
The edge modes are marked by red solid lines. Red lines are two-fold degenerate.} 
\label{fig-edge}
\end{figure}

We use the mean-field approximation by replacing the bosonic operator $b_j$ by $b_j\approx \langle b_j \rangle=B$ \cite{lattice-crossover}.
According to the Bardeen-Cooper-Schrieffer theory,
it is easy to see $B$ characterizes the order parameter for the superfluid phase.
The details of the mean-field approach are presented in Appendix \ref{sec-app-mf}.
FIG. \ref{fig-edge}(a) shows the phase diagram in the $\eta$-$n$ plane at zero temperature.
There exist three superfluid phases in the diagram.
By changing the filling factor $n$,
the band gap of the trivial superfluid phase closes when the chemical potential $\mu$ equals $\mu_{c1}=\mathcal{E}_0\pm \mathcal{E}_1$, transitioning to a gapless superfluid state,
and reopens at $\mu_{c2}=\mathcal{E}_0\pm \mathcal{E}_2$, which corresponds to a topological superfluid region.
Here $\mathcal{E}_0=\eta^2JU_0/U_1$, $\mathcal{E}_1=2\eta\alpha(2-\eta^2 U_0/8U_1)^{1/2}$,
and $\mathcal{E}_2=2\eta\alpha(\eta U_0/U_1-\eta^2U_0^2/4U_1^2)^{1/2}$.
Since the system respects the particle-hole as well as time-reversal symmetries,
the topological superfluid phase supports four-fold degenerate MZMs, as shown in FIG. \ref{fig-edge}(b),
and they are localized on edges of the square bulk \cite{time-sc-Zhang}.

For 2D Fermi gases at nonzero temperature,
the phase fluctuations of the order parameters play the essential role that suppresses superfluidity \cite{bkt1,bkt2,bkt3,bkt4,bkt5}.
The system will undergo a transition to the normal phase when the temperature exceeds a critical value 
that is known as the Berezinskii-Kosterlitz-Thouless transition temperature \cite{bkt-origin-1,bkt-origin-2,bkt-origin-3}.
In order to investigate the stability of superfluid phases against fluctuations,
we plot the phase diagram at non-zero temperature in Fig. \ref{fig-Tc}.
The detailed formulas by accounting for the phase fluctuations are given in Appendix \ref{sec-app-bkt}.
From Fig. \ref{fig-Tc}, we find that all the superfluid phases, including the topological one,
are robust even though the fluctuations are present.
The critical temperature of superfluid phases (i.e., the boundary between superfluid phases and the normal gas)
slightly decreases with increasing $\eta$.

\begin{figure}[t]
\centering
\includegraphics[width=0.4\textwidth]{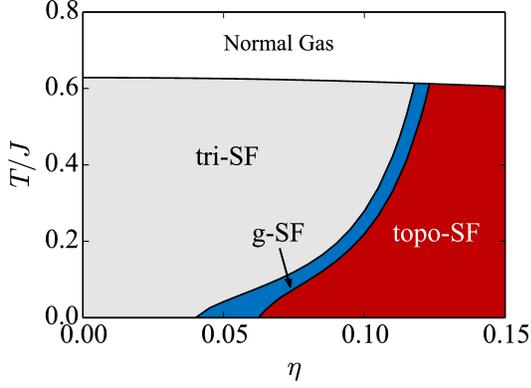}
\caption{Phase diagram at nonzero temperature $T$.
We set $n=1.6$, $U_0=5.0J$, $U_1=0.32J$, $\nu_0=1.0J$, and $\alpha=0.5J$.} 
\label{fig-Tc}
\end{figure}

\subsection{Corner MZM}
The previous example focuses on a single-layer system.
In the multilayer one, for simplicity a bilayer lattice system,
the corner MZMs have been actively investigated in recent research \cite{zhang-corner,wang-corner},
the most important feature of which is that their wave function is localized on the corner of the square lattice.
Based on the recent investigations, the lattice systems that support the corner MZMs have the following two similarities:
(i) The single-particle Hamiltonian is a topological insulator,
for instance, the quantum spin Hall insulator \cite{2d-topo-ins}.
(ii) The interacting Hamiltonian is composed of tunable on-site and off-site components.

We consider the following bilayer Hamiltonian in company with SOC,
\begin{align}
H_{\rm BL} &= \sum_{n} [H_{\rm hop}^{(n)} +H_Z^{(n)}+ H_{\rm soc}^{(n)}] + H_B +H_{\rm int} \,,\\
H_{\rm hop}^{(n)} &= -\sum_{j,l}\sum_{\tau\tau'} J_l c_{jn\tau}^\dag [s_z]_{\tau\tau'} c_{j+{\bf e}_l, n\tau'} + \text{H.c.} \label{eq-corner-hop}\\
H_Z^{(n)} &= \sum_{j,\tau,\tau'} m_0 c_{jn\tau}^\dag [s_z]_{\tau\tau'} c_{jn\tau'}
-\sum_{j,\tau} \mu c_{jn\tau}^\dag c_{jn\tau} \,,\\
H_B &=- \sum_{\langle ij \rangle} J_B b_{i}^\dag b_{j}-\sum_{j}\mu_B b_{i}^\dag b_{j} \,,\\
H_{\rm soc}^{(n)} &= i\alpha \sum_{j,\tau,\tau'}
(c_{jn\tau}^\dag [s_x]_{\tau\tau'} c_{j+{\bf e}_y, n\tau'} \notag\\
&\quad - c_{jn\tau}^\dag [s_y]_{\tau\tau'} c_{j+{\bf e}_x, n\tau'}) +\text{H.c.} \label{eq-corner-soc}
\end{align}
Here $n=1,2$ denotes the layer index, and $m_0$ characterizes the spin imbalance.
We can see the nearest-neighbor hopping of spin-$\uparrow\downarrow$ atoms hosts opposite signs.

The interacting Hamiltonian in the two-channel model is given by
\begin{align}
&H_\mathrm{int} = \sum_{j,l} [\sum_n U_0' b_{j}^\dag c_{jn\uparrow} c_{jn\downarrow}
+U_0 b_{j}^\dag(c_{j1\uparrow} c_{j2\downarrow} + c_{j2\uparrow} c_{j1\downarrow}) \notag\\
&+\frac{U_1}{2} (b_{j}^\dag+b_{j+{\bf e}_l}^\dag)(c_{j1\uparrow} c_{j+{\bf e}_l2\downarrow} + c_{j2\uparrow} c_{j+{\bf e}_l1\downarrow})] +\text{H.c.} \label{eq-corner-h-int-1}
\end{align}
Here the $U_0'$ term stems from the intra-layer interaction,
while the $U_0$ and $U_1$ terms are from the inter-layer one.
We prepare the bosonic molecule states trapped in the center of two layers of fermions,
and thus simultaneously take the intra- and inter-layer interaction into consideration.
We then impose the periodical driving term $H_\mathrm{dr} = \sum_{j,n,\tau}V_\mathrm{dr}'(t)c_{jn\tau}^\dag c_{jn\tau}$,
where $V_\mathrm{dr}'(t)$ is generated by adding a layer-index-dependent term to $V_{\rm dr}(t)$ of Eq.(\ref{eq-driven-field}),
\begin{equation}
V_\mathrm{dr}'(t) = V_\mathrm{dr}(t) + (-1)^n \omega' \,.\label{eq-corner-driven-field}
\end{equation}
We repeat the rotation transformation (see Eq.(\ref{eq-rotate})). 
For simplicity, we choose $\omega'\approx 1.3\omega_{\rm dr}$ in Eq.(\ref{eq-corner-driven-field}).
Thus in the rotating frame,
the $\omega'$ term has no influence on the inter-layer interaction in Eq.(\ref{eq-corner-h-int-1}).
By contrast, the intra-layer interaction $U_0'$ term in Eq.(\ref{eq-corner-h-int-1}) will be rotated off.
After neglecting the rapidly oscillating terms, we obtain the effective Hamiltonian as
\begin{equation}
\mathcal{H}_\mathrm{BL}=\sum_{n}[\eta H_\mathrm{hop}^{(n)}+H_{Z}^{(n)}+\eta H_\mathrm{soc}^{(n)}] +H_B +\mathcal{H}_\mathrm{int} \,,\label{eq-corner-h-final}
\end{equation}
where
\begin{align}
&\mathcal{H}_\mathrm{int} = 
\sum_{j}\eta U_0 b_j^\dag (c_{j1\uparrow} c_{j2\downarrow} + c_{j2\uparrow} c_{j1\downarrow}) \notag\\
& +\frac{U_1}{2} (b_{j}^\dag+b_{j+{\bf e}_l}^\dag) (c_{j1\uparrow} c_{j+{\bf e}_l2\downarrow} + c_{j2\uparrow} c_{j+{\bf e}_l1\downarrow}) +\text{H.c.} \label{eq-corner-h-int-2}
\end{align}
In Eq. (\ref{eq-corner-h-int-2}) we can obtain the similar conclusion that on-site inter-layer interaction ($\eta U_0$) and the off-site one ($U_1$) are individually tunable by the driving field.

\begin{figure}[t]
\centering
\includegraphics[width=0.5\textwidth]{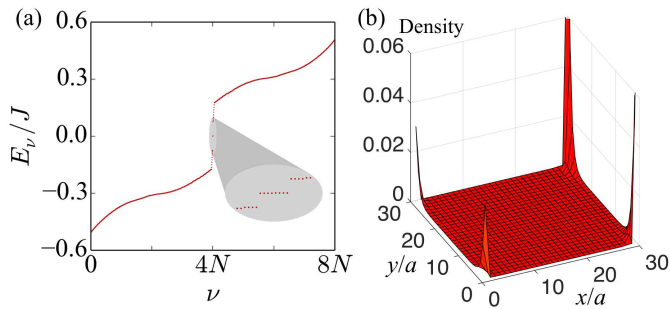}
\caption{
(a) BdG spectrum of the lattice system with the open boundary condition in both $x$ and $y$ directions.
The inset (the gray region) shows the magnified illustration of the zero-energy vicinity, in which eight zero-energy states exist in the center of the band gap.
(b) Spatial distribution of one corner MZM.
We calculate a 2D lattice with $N=L\times L$ sites and set $L=30$.
Other parameters are $U_0=5.0J$, $U_1=0.32J$, $\mu=0.0$, $\nu_0=1.0J$, $\alpha=1.0J$,
$\eta=0.1$, and $J_x=-J_y=J$.} 
\label{fig-corner}
\end{figure}

We repeat the numeric mean-field approach, and show the results in FIG. \ref{fig-corner}.
The Hamiltonian (\ref{eq-corner-h-final}) preserves the particle-hole symmetry,
and remains unchanged if one exchanges the layer index.
Furthermore, it is invariant under a rotation by an angle $2\pi$ along the $z$ axis of spin space associated with a mirror reflection in real space.
Therefore the BdG spectrum is eight-fold degenerate, as shown in FIG. \ref{fig-corner}(a).
We plot the spatial distribution of one zero mode in FIG. \ref{fig-corner}(b),
and clearly see that the wave function of the zero mode is dramatically localized on four corners of the square bulk,
yielding the emergence of corner MZMs.

\section{Experimental implementation}\label{sec-exp}

In experiments,
the driving field $V_\mathrm{dr}(t)$ can be introduced by imprinting two pairs of counter-propagating lasers on the atoms.
One pair drives the transition from the pseudo-spin states to excited states.
The adiabatic elimination of the excited states
generates a time-dependent ac-Stark shift for the atoms:
$\Gamma\cos(\omega_\mathrm{dr} t)\cos(k_\mathrm{dr}x+k_\mathrm{dr}y)$ with $\Gamma$ as its magnitude.
The shift exhibits a standing-wave mode.
The other pair creates a time-independent shift
$\Gamma'\cos(k_\mathrm{dr}x+k_\mathrm{dr}y)$
the magnitude of which is equal to the driving frequency: $\Gamma'=\omega_\mathrm{dr}$.
When we tune $k_\mathrm{dr}$ equal to the optical lattice wavevector $k_L$,
the spatial distribution of the Stark shifts will exhibit a checkerboard structure $(-1)^{j_x+j_y}$ with respect to the site index $j$.
We remark that by choosing proper excited states, the other levels' (including the bosonic molecular states) transitions are far detuned.
In this way, it is attainable that the driving field does not act on the bosonic molecular states.

The engineering of edge MZMs is readily realized in current cold-atom techniques,
since SOC has been successfully realized via Raman protocols \cite{2d-soc-exp-1,2d-soc-exp-2}.
For corner MZMs, the SOC terms (\ref{eq-corner-soc}) of the two layers have opposite sign.
This can be realized if the strength of lasers that generate SOC is designed to be spatially modulated along the normal direction of the lattice plane,
resulting in $\alpha(z)=\alpha\cos(k_Lz)$.
For engineering the hopping term (\ref{eq-corner-hop}),
we can use the laser-assisted hopping technique \cite{laser-hop-1,laser-hop-2} to generate hopping 
accompanied by a $\pi$-phase difference not only between opposite spins but also between $x$ and $y$ directions.
Thus a spin-dependent hopping can be obtained.

In ultracold Fermi gases, collisional heating from periodic driving is suppressed due to the Pauli blocking of atomic collisions at low temperature \cite{pauli-blocking}.
Instead, the absorption of photons from the driving field plays the key role for the heating effect,
in which the heating rate is proportional to the driving amplitude $\Gamma$, however is independent from the driving frequency $\omega_{\rm dr}$ \cite{floquet-heating}.
In our proposal, we prepare the driving field in the weak $\eta$ regime (i.e., $\Gamma\ll\hbar\omega$) to suppress the on-site interaction.
Therefore, the heating effect can be reduced in the rapidly driving limit with a weak amplitude.

\section{Conclusion}\label{sec-conclude}

In summary, we present a valid and feasible proposal for engineering the effective Hamiltonian in company with tunable interaction based on driven optical lattices.
Our proposal hosts the following two features:
(i) the mixed pairing of the effective Hamiltonian is individually tunable via the driving fields rather than magnetic Feshbach resonances;
and (ii) it can be applied in engineering MZMs localized in edges (resp. corners) of the 2D lattice system in the single-layer (resp. bilayer) scheme.
Therefore, the proposal offers a potential candidate for engineering and studying topological superfluids supporting MZMs in ultracold atoms.

\section{Acknowledgements}
This work is supported by
National Natural Science Foundation of China (Grants No. 11474271, 11674305, and 11704367).

\appendix

\section{Mean-field Approach}\label{sec-app-mf}

For Hamiltonian (\ref{eq-h-app-edge}),
we use the mean-field approximation by assuming $b_j\approx \langle b_j \rangle= B$,
and exploit the BdG transformation
\begin{equation}
c_{j\sigma} = \sum_{\nu=1}^{4N} (u^\nu_{\phi(j),\sigma} \gamma_\nu + v^\nu_{\phi(j),\sigma} \gamma_\nu^\dag) \,,
\end{equation}
where $\gamma$'s are the quasi-particle operators,
$N=L\times L$ ($L$ is the length of the square bulk),
and $\phi(j)=j_x+(j_y-1)L=1,\cdots,N$ is the mapped index for the original 2D lattice's $j$-th site.
The $\gamma$'s coefficients $\hat{u}_\nu=(u_{1\uparrow}^\nu,\cdots,u_{N\uparrow}^\nu,u_{1\downarrow}^\nu,\cdots,u_{N\downarrow}^\nu)$ 
and $\hat{v}_\nu=(v_{1\downarrow}^\nu,\cdots,v_{N\downarrow}^\nu,-v_{1\uparrow}^\nu,\cdots,-v_{N\uparrow}^\nu)$ 
satisfy the following equations (we denote $\Psi=(\hat{u}_\nu,\hat{v}_\nu)^T$)
\begin{equation}
[\tau_z\otimes(\hat{D}-\hat{X}s_x+\hat{Y}s_y)+\tau_x\otimes\hat{B} ]\Psi
=E_\nu \Psi \,.
\end{equation}

Here $E_\nu$ gives the BdG spectrum,
and $\tau_{x,y,z}$ are Pauli matrices in the particle-hole space.
$\hat{D}$, $\hat{X}$, $\hat{Y}$, and $\hat{B}$ are $N\times N$ matrices the elements of which are given as follows:
\begin{equation}
\left\{\begin{split}
&\hat{D}_{\phi(i)\phi(j)} = -\mu\delta_{ij} - \eta J(\delta_{i-j,{\bf e}_{x,y}}+\delta_{j-i,{\bf e}_{x,y}}) \\
&\hat{X}_{\phi(i)\phi(j)} = i\eta\alpha (\delta_{i-j,{\bf e}_{y}}-\delta_{j-i,{\bf e}_{y}}) \\
&\hat{Y}_{\phi(i)\phi(j)} = -i\eta\alpha (\delta_{i-j,{\bf e}_{x}}-\delta_{j-i,{\bf e}_{x}}) \\
&\hat{B}_{\phi(i)\phi(j)} = \eta U_0 B \delta_{ij} + U_1 B (\delta_{i-j,{\bf e}_{x,y}}+\delta_{j-i,{\bf e}_{x,y}})
\end{split}\right.\,.\notag
\end{equation}
Here $\delta_{ij}$ is the Kronecker-$\delta$ function.
The Hamiltonian (12) of the maintext is thus cast into a quadratic form
\begin{equation}
\mathcal{H}_\mathrm{2D} = \sum_\nu (E_\nu \gamma_\nu^\dag \gamma_\nu -\frac{1}{2}) + \epsilon_0 \,,
\end{equation}
where $\epsilon_0=(\nu_0-2\mu)|B|^2$ is the boson's energy.
The system energy is thereby given by \cite{xu-mf}
\begin{equation}
\mathcal{E}=\langle \mathcal{H}\rangle = \sum_\nu E_\nu[ f(E_\nu)-\sum_{j,\sigma} |v_{\phi(j),\sigma}^\nu|^2 ]+\epsilon_0 
\,,\label{eq-edge-energy-eq}
\end{equation}
where $f(\cdot)$ represents the Fermi distribution at temperature $T$.
The number equation is then expressed as
\begin{equation}
n= \sum_\nu \langle \gamma_\nu^\dag \gamma_\nu \rangle
=\sum_{\nu,j,\sigma}  |u_{\phi(j),\sigma}^\nu|^2f(E_\nu) + |v_{\phi(j),\sigma}^\nu|^2f(-E_\nu)
\,. \label{eq-edge-number-eq}
\end{equation}
Here $n$ is the filling factor per site.
Under the number conservation constraint Eq.(\ref{eq-edge-number-eq}), 
we can obtain the order parameter $B$ and the chemical potential $\mu$ by self-consistently minimizing the system energy Eq. (\ref{eq-edge-energy-eq}) with respect to $B$.

\section{Phase Fluctuations}\label{sec-app-bkt}

The partition function of the 2D Fermi gas described by Hamiltonian (\ref{eq-h-app-edge}) is
\begin{equation}
\mathcal{Z}=\int \mathcal{D}\psi \,e^{-S_{\text{eff}}[\psi]}\,,
\end{equation}
where $\beta=\frac{1}{T}$ at temperature $T$, and the effective action can be expressed as
\begin{equation}
S_\text{eff}[\psi]=\int d \tau d {\bf r} \sum_\sigma \psi^\ast_\sigma({\bf r},\tau) \partial_\tau \psi_\sigma({\bf r},\tau) + \mathcal{H}_{\rm 2D} \,. \label{eq-append-s-1}
\end{equation}
We use the mean-field approximation of the order parameter (see Appendix \ref{sec-app-mf}) and integrate out the $\psi_\sigma$ fields.
Under the basis $\Psi({\bf k})=(\psi_{{\bf k},\uparrow}, \psi_{{\bf k},\downarrow}, \psi^\dag_{-{\bf k},\downarrow}, -\psi^\dag_{-{\bf k},\uparrow})^T$
with ${\bf k}\equiv -i\nabla$, the effective action (\ref{eq-append-s-1}) is rewritten as
\begin{equation}
S_\text{eff}=\int d \tau d {\bf r} ( \epsilon_0 -\frac{1}{2}\text{Trln}G^{-1} ) \label{eq-append-s-2}
\end{equation}
where the inverse Green's function $G^{-1}$ is expressed as
\begin{equation}
G^{-1}= -\partial_\tau - H_{\rm BdG} \label{eq-append-green}
\end{equation}
and the BdG Hamiltonian is written as
\begin{equation}
H_{\rm BdG} =
\begin{pmatrix}
H_0({\bf k}) & B \\
B^\dag & s_y H_0^*(-{\bf k})s_y
\end{pmatrix} \,.\label{eq-append-green-bdg}
\end{equation}
Here the single-particle term $H_0({\bf k})=\xi_{\bf k}+\eta H_{\rm soc}$ with $\xi_{\bf k}=\eta k^2/2m-\mu$.
$\epsilon_0=\sum_{{\bf k}\sigma}\xi_{\bf k}/2$ is introduced due to the anti-commutation of $\psi_\sigma$ fields.

In 2D Fermi gases,
the phase fluctuation of the order parameter plays the essential role in the superfluid phase transition.
It can be introduced by imposing a perturbative phase $\theta$ into the order parameter $B$, i.e., $B\rightarrow Be^{i\theta}$ \cite{bkt5}.
Under the unitary rotation $\hat{U}=\exp(i\theta/2)\,\tau_z\otimes\mathbb{I}$,
the inverse Green's function is given by the following form composed by two parts,
\begin{equation}
\widetilde{G}^{-1}(\theta)=\hat{U}^\dag G^{-1} \hat{U}=G^{-1}-\Sigma(\theta)
\end{equation}
The first item is the original $\theta$-independent form Eq.(\ref{eq-append-green}),
while the second term $\Sigma$ is the $\theta$-dependent self energy expressed as
\begin{align}
\Sigma(\theta)&=\Big(\frac{i}{2}\partial_\tau \theta+\frac{\eta(\nabla\theta)^2}{8m}\Big)\tau_z\otimes\mathbb{I} \notag\\
&-\Big(\frac{i\eta{\nabla}^2\theta}{4m}+\frac{i\eta\nabla\theta\cdot\nabla}{2m}\Big)\mathbb{I}\otimes\mathbb{I} \notag\\
&+\frac{\eta\alpha}{2}(\partial_x\theta\,\mathbb{I}\otimes s_y-\partial_y\theta\,\mathbb{I}\otimes s_x) \,.
\end{align}
Correspondingly, the effective action (\ref{eq-append-s-2}) is given by
\begin{equation}
S_\text{eff}=S_\text{mf}+S_\text{fluc}
\end{equation}
with the mean-field term
\begin{equation}
S_\text{mf}=\int d \tau d {\bf r} ( \epsilon_0-\frac{1}{2}\text{Trln}G^{-1} )
\end{equation}
and the fluctuation induced term
\begin{align}
S_\text{fluc}&=-\frac{1}{2}\int d \tau d {\bf r}\, \text{Tr}\ln(1-G\Sigma) \\
&\approx\frac{1}{2}\int d \tau d {\bf r}\, \text{Tr}(G\Sigma+G\Sigma G\Sigma/2) \\
&\equiv\frac{1}{2}\int d {\bf r} d \tau\, [J_{xx}(\partial_x\theta)^2+J_{yy}(\partial_y\theta)^2\notag\\
&+J_{xy}\partial_x\theta\partial_y\theta+P(\partial_\tau\theta)^2-i A\partial_\tau\theta] \label{eq-append-s-fluc}
\end{align}
The critical temperature $T_c$ of the superfluid phase transition can be determined by \cite{bkt3,bkt4,bkt6}
\begin{equation}
T_c = \frac{\pi}{2} \sqrt{J_{xx}J_{yy}} \,,
\end{equation}
where $J_{xx}$ and $J_{yy}$ can be obtained from Eq.(\ref{eq-append-s-fluc}):
\begin{equation}
J_{xx(yy)}=\frac{n}{4m}+\frac{\beta}{8}\sum_{{\bf k},\zeta}(\frac{\eta^2 k_{x(y)}^2}{m^2}+\eta^2\alpha^2)f(E_\zeta)[f(E_\zeta)-1] \,.
\end{equation}
Here $E_{\zeta}$ ($\zeta=1\sim4$) are the eigenvalues of BdG Hamiltonian (\ref{eq-append-green-bdg}).

\bibliographystyle{apsrev4-1}
\bibliography{bib}

\begin{thebibliography}{73}%
\makeatletter
\providecommand \@ifxundefined [1]{%
 \@ifx{#1\undefined}
}%
\providecommand \@ifnum [1]{%
 \ifnum #1\expandafter \@firstoftwo
 \else \expandafter \@secondoftwo
 \fi
}%
\providecommand \@ifx [1]{%
 \ifx #1\expandafter \@firstoftwo
 \else \expandafter \@secondoftwo
 \fi
}%
\providecommand \natexlab [1]{#1}%
\providecommand \enquote  [1]{``#1''}%
\providecommand \bibnamefont  [1]{#1}%
\providecommand \bibfnamefont [1]{#1}%
\providecommand \citenamefont [1]{#1}%
\providecommand \href@noop [0]{\@secondoftwo}%
\providecommand \href [0]{\begingroup \@sanitize@url \@href}%
\providecommand \@href[1]{\@@startlink{#1}\@@href}%
\providecommand \@@href[1]{\endgroup#1\@@endlink}%
\providecommand \@sanitize@url [0]{\catcode `\\12\catcode `\$12\catcode
  `\&12\catcode `\#12\catcode `\^12\catcode `\_12\catcode `\%12\relax}%
\providecommand \@@startlink[1]{}%
\providecommand \@@endlink[0]{}%
\providecommand \url  [0]{\begingroup\@sanitize@url \@url }%
\providecommand \@url [1]{\endgroup\@href {#1}{\urlprefix }}%
\providecommand \urlprefix  [0]{URL }%
\providecommand \Eprint [0]{\href }%
\providecommand \doibase [0]{http://dx.doi.org/}%
\providecommand \selectlanguage [0]{\@gobble}%
\providecommand \bibinfo  [0]{\@secondoftwo}%
\providecommand \bibfield  [0]{\@secondoftwo}%
\providecommand \translation [1]{[#1]}%
\providecommand \BibitemOpen [0]{}%
\providecommand \bibitemStop [0]{}%
\providecommand \bibitemNoStop [0]{.\EOS\space}%
\providecommand \EOS [0]{\spacefactor3000\relax}%
\providecommand \BibitemShut  [1]{\csname bibitem#1\endcsname}%
\let\auto@bib@innerbib\@empty
\bibitem [{\citenamefont {Nayak}\ \emph {et~al.}(2008)\citenamefont {Nayak},
  \citenamefont {Simon}, \citenamefont {Stern}, \citenamefont {Freedman},\ and\
  \citenamefont {Das~Sarma}}]{quant-compu-rev}%
  \BibitemOpen
  \bibfield  {author} {\bibinfo {author} {\bibfnamefont {C.}~\bibnamefont
  {Nayak}}, \bibinfo {author} {\bibfnamefont {S.~H.}\ \bibnamefont {Simon}},
  \bibinfo {author} {\bibfnamefont {A.}~\bibnamefont {Stern}}, \bibinfo
  {author} {\bibfnamefont {M.}~\bibnamefont {Freedman}}, \ and\ \bibinfo
  {author} {\bibfnamefont {S.}~\bibnamefont {Das~Sarma}},\ }\href {\doibase
  10.1103/RevModPhys.80.1083} {\bibfield  {journal} {\bibinfo  {journal} {Rev.
  Mod. Phys.}\ }\textbf {\bibinfo {volume} {80}},\ \bibinfo {pages} {1083}
  (\bibinfo {year} {2008})}\BibitemShut {NoStop}%
\bibitem [{\citenamefont {Hasan}\ and\ \citenamefont
  {Kane}(2010)}]{topo-cri-rev-1}%
  \BibitemOpen
  \bibfield  {author} {\bibinfo {author} {\bibfnamefont {M.~Z.}\ \bibnamefont
  {Hasan}}\ and\ \bibinfo {author} {\bibfnamefont {C.~L.}\ \bibnamefont
  {Kane}},\ }\href {\doibase 10.1103/RevModPhys.82.3045} {\bibfield  {journal}
  {\bibinfo  {journal} {Rev. Mod. Phys.}\ }\textbf {\bibinfo {volume} {82}},\
  \bibinfo {pages} {3045} (\bibinfo {year} {2010})}\BibitemShut {NoStop}%
\bibitem [{\citenamefont {Qi}\ and\ \citenamefont
  {Zhang}(2011)}]{topo-cri-rev-2}%
  \BibitemOpen
  \bibfield  {author} {\bibinfo {author} {\bibfnamefont {X.-L.}\ \bibnamefont
  {Qi}}\ and\ \bibinfo {author} {\bibfnamefont {S.-C.}\ \bibnamefont {Zhang}},\
  }\href {\doibase 10.1103/RevModPhys.83.1057} {\bibfield  {journal} {\bibinfo
  {journal} {Rev. Mod. Phys.}\ }\textbf {\bibinfo {volume} {83}},\ \bibinfo
  {pages} {1057} (\bibinfo {year} {2011})}\BibitemShut {NoStop}%
\bibitem [{\citenamefont {Kitaev}(2001)}]{kitaev-p-wave}%
  \BibitemOpen
  \bibfield  {author} {\bibinfo {author} {\bibfnamefont {A.~Y.}\ \bibnamefont
  {Kitaev}},\ }\href {http://stacks.iop.org/1063-7869/44/i=10S/a=S29}
  {\bibfield  {journal} {\bibinfo  {journal} {Physics-Uspekhi}\ }\textbf
  {\bibinfo {volume} {44}},\ \bibinfo {pages} {131} (\bibinfo {year}
  {2001})}\BibitemShut {NoStop}%
\bibitem [{\citenamefont {Zhang}\ \emph {et~al.}(2008)\citenamefont {Zhang},
  \citenamefont {Tewari}, \citenamefont {Lutchyn},\ and\ \citenamefont
  {Das~Sarma}}]{soc-p-wave}%
  \BibitemOpen
  \bibfield  {author} {\bibinfo {author} {\bibfnamefont {C.}~\bibnamefont
  {Zhang}}, \bibinfo {author} {\bibfnamefont {S.}~\bibnamefont {Tewari}},
  \bibinfo {author} {\bibfnamefont {R.~M.}\ \bibnamefont {Lutchyn}}, \ and\
  \bibinfo {author} {\bibfnamefont {S.}~\bibnamefont {Das~Sarma}},\ }\href
  {\doibase 10.1103/PhysRevLett.101.160401} {\bibfield  {journal} {\bibinfo
  {journal} {Phys. Rev. Lett.}\ }\textbf {\bibinfo {volume} {101}},\ \bibinfo
  {pages} {160401} (\bibinfo {year} {2008})}\BibitemShut {NoStop}%
\bibitem [{\citenamefont {Deng}\ \emph {et~al.}(2012)\citenamefont {Deng},
  \citenamefont {Viola},\ and\ \citenamefont {Ortiz}}]{time-sc-Deng}%
  \BibitemOpen
  \bibfield  {author} {\bibinfo {author} {\bibfnamefont {S.}~\bibnamefont
  {Deng}}, \bibinfo {author} {\bibfnamefont {L.}~\bibnamefont {Viola}}, \ and\
  \bibinfo {author} {\bibfnamefont {G.}~\bibnamefont {Ortiz}},\ }\href
  {\doibase 10.1103/PhysRevLett.108.036803} {\bibfield  {journal} {\bibinfo
  {journal} {Phys. Rev. Lett.}\ }\textbf {\bibinfo {volume} {108}},\ \bibinfo
  {pages} {036803} (\bibinfo {year} {2012})}\BibitemShut {NoStop}%
\bibitem [{\citenamefont {Zhang}\ \emph {et~al.}(2013)\citenamefont {Zhang},
  \citenamefont {Kane},\ and\ \citenamefont {Mele}}]{time-sc-Zhang}%
  \BibitemOpen
  \bibfield  {author} {\bibinfo {author} {\bibfnamefont {F.}~\bibnamefont
  {Zhang}}, \bibinfo {author} {\bibfnamefont {C.~L.}\ \bibnamefont {Kane}}, \
  and\ \bibinfo {author} {\bibfnamefont {E.~J.}\ \bibnamefont {Mele}},\ }\href
  {\doibase 10.1103/PhysRevLett.111.056402} {\bibfield  {journal} {\bibinfo
  {journal} {Phys. Rev. Lett.}\ }\textbf {\bibinfo {volume} {111}},\ \bibinfo
  {pages} {056402} (\bibinfo {year} {2013})}\BibitemShut {NoStop}%
\bibitem [{\citenamefont {Keselman}\ \emph {et~al.}(2013)\citenamefont
  {Keselman}, \citenamefont {Fu}, \citenamefont {Stern},\ and\ \citenamefont
  {Berg}}]{time-sc-Keselman}%
  \BibitemOpen
  \bibfield  {author} {\bibinfo {author} {\bibfnamefont {A.}~\bibnamefont
  {Keselman}}, \bibinfo {author} {\bibfnamefont {L.}~\bibnamefont {Fu}},
  \bibinfo {author} {\bibfnamefont {A.}~\bibnamefont {Stern}}, \ and\ \bibinfo
  {author} {\bibfnamefont {E.}~\bibnamefont {Berg}},\ }\href {\doibase
  10.1103/PhysRevLett.111.116402} {\bibfield  {journal} {\bibinfo  {journal}
  {Phys. Rev. Lett.}\ }\textbf {\bibinfo {volume} {111}},\ \bibinfo {pages}
  {116402} (\bibinfo {year} {2013})}\BibitemShut {NoStop}%
\bibitem [{\citenamefont {Wang}\ \emph
  {et~al.}(2018{\natexlab{a}})\citenamefont {Wang}, \citenamefont {Liu},
  \citenamefont {Lu},\ and\ \citenamefont {Zhang}}]{zhang-corner}%
  \BibitemOpen
  \bibfield  {author} {\bibinfo {author} {\bibfnamefont {Q.}~\bibnamefont
  {Wang}}, \bibinfo {author} {\bibfnamefont {C.-C.}\ \bibnamefont {Liu}},
  \bibinfo {author} {\bibfnamefont {Y.-M.}\ \bibnamefont {Lu}}, \ and\ \bibinfo
  {author} {\bibfnamefont {F.}~\bibnamefont {Zhang}},\ }\href {\doibase
  10.1103/PhysRevLett.121.186801} {\bibfield  {journal} {\bibinfo  {journal}
  {Phys. Rev. Lett.}\ }\textbf {\bibinfo {volume} {121}},\ \bibinfo {pages}
  {186801} (\bibinfo {year} {2018}{\natexlab{a}})}\BibitemShut {NoStop}%
\bibitem [{\citenamefont {Yan}\ \emph {et~al.}(2018)\citenamefont {Yan},
  \citenamefont {Song},\ and\ \citenamefont {Wang}}]{wang-corner}%
  \BibitemOpen
  \bibfield  {author} {\bibinfo {author} {\bibfnamefont {Z.}~\bibnamefont
  {Yan}}, \bibinfo {author} {\bibfnamefont {F.}~\bibnamefont {Song}}, \ and\
  \bibinfo {author} {\bibfnamefont {Z.}~\bibnamefont {Wang}},\ }\href {\doibase
  10.1103/PhysRevLett.121.096803} {\bibfield  {journal} {\bibinfo  {journal}
  {Phys. Rev. Lett.}\ }\textbf {\bibinfo {volume} {121}},\ \bibinfo {pages}
  {096803} (\bibinfo {year} {2018})}\BibitemShut {NoStop}%
\bibitem [{\citenamefont {Zhu}(2018)}]{zhu-corner-2018}%
  \BibitemOpen
  \bibfield  {author} {\bibinfo {author} {\bibfnamefont {X.}~\bibnamefont
  {Zhu}},\ }\href {\doibase 10.1103/PhysRevB.97.205134} {\bibfield  {journal}
  {\bibinfo  {journal} {Phys. Rev. B}\ }\textbf {\bibinfo {volume} {97}},\
  \bibinfo {pages} {205134} (\bibinfo {year} {2018})}\BibitemShut {NoStop}%
\bibitem [{\citenamefont {Liu}\ \emph {et~al.}(2018)\citenamefont {Liu},
  \citenamefont {He},\ and\ \citenamefont {Nori}}]{liu-corner}%
  \BibitemOpen
  \bibfield  {author} {\bibinfo {author} {\bibfnamefont {T.}~\bibnamefont
  {Liu}}, \bibinfo {author} {\bibfnamefont {J.~J.}\ \bibnamefont {He}}, \ and\
  \bibinfo {author} {\bibfnamefont {F.}~\bibnamefont {Nori}},\ }\href {\doibase
  10.1103/PhysRevB.98.245413} {\bibfield  {journal} {\bibinfo  {journal} {Phys.
  Rev. B}\ }\textbf {\bibinfo {volume} {98}},\ \bibinfo {pages} {245413}
  (\bibinfo {year} {2018})}\BibitemShut {NoStop}%
\bibitem [{\citenamefont {Huang}\ and\ \citenamefont
  {Liu}(2018)}]{vincent-corner}%
  \BibitemOpen
  \bibfield  {author} {\bibinfo {author} {\bibfnamefont {B.}~\bibnamefont
  {Huang}}\ and\ \bibinfo {author} {\bibfnamefont {W.~V.}\ \bibnamefont
  {Liu}},\ }\href@noop {} {\bibfield  {journal} {\bibinfo  {journal} {arXiv
  preprint arXiv:1811.00555}\ } (\bibinfo {year} {2018})}\BibitemShut {NoStop}%
\bibitem [{\citenamefont {Zhu}(2019)}]{zhu-corner-2019}%
  \BibitemOpen
  \bibfield  {author} {\bibinfo {author} {\bibfnamefont {X.}~\bibnamefont
  {Zhu}},\ }\href {\doibase 10.1103/PhysRevLett.122.236401} {\bibfield
  {journal} {\bibinfo  {journal} {Phys. Rev. Lett.}\ }\textbf {\bibinfo
  {volume} {122}},\ \bibinfo {pages} {236401} (\bibinfo {year}
  {2019})}\BibitemShut {NoStop}%
\bibitem [{\citenamefont {Luo}\ and\ \citenamefont
  {Zhang}(2019)}]{luo-corner-2019}%
  \BibitemOpen
  \bibfield  {author} {\bibinfo {author} {\bibfnamefont {X.-W.}\ \bibnamefont
  {Luo}}\ and\ \bibinfo {author} {\bibfnamefont {C.}~\bibnamefont {Zhang}},\
  }\href {\doibase 10.1103/PhysRevLett.123.073601} {\bibfield  {journal}
  {\bibinfo  {journal} {Phys. Rev. Lett.}\ }\textbf {\bibinfo {volume} {123}},\
  \bibinfo {pages} {073601} (\bibinfo {year} {2019})}\BibitemShut {NoStop}%
\bibitem [{\citenamefont {Yan}(2019)}]{yan-corner-2019}%
  \BibitemOpen
  \bibfield  {author} {\bibinfo {author} {\bibfnamefont {Z.}~\bibnamefont
  {Yan}},\ }\href {\doibase 10.1103/PhysRevLett.123.177001} {\bibfield
  {journal} {\bibinfo  {journal} {Phys. Rev. Lett.}\ }\textbf {\bibinfo
  {volume} {123}},\ \bibinfo {pages} {177001} (\bibinfo {year}
  {2019})}\BibitemShut {NoStop}%
\bibitem [{\citenamefont {Bloch}\ \emph {et~al.}(2012)\citenamefont {Bloch},
  \citenamefont {Dalibard},\ and\ \citenamefont
  {Nascimbene}}]{quantum-simu-rev-nphys}%
  \BibitemOpen
  \bibfield  {author} {\bibinfo {author} {\bibfnamefont {I.}~\bibnamefont
  {Bloch}}, \bibinfo {author} {\bibfnamefont {J.}~\bibnamefont {Dalibard}}, \
  and\ \bibinfo {author} {\bibfnamefont {S.}~\bibnamefont {Nascimbene}},\
  }\href {\doibase doi:10.1038/nphys2259} {\bibfield  {journal} {\bibinfo
  {journal} {Nat. Phys.}\ }\textbf {\bibinfo {volume} {8}},\ \bibinfo {pages}
  {267} (\bibinfo {year} {2012})}\BibitemShut {NoStop}%
\bibitem [{\citenamefont {Gross}\ and\ \citenamefont
  {Bloch}(2017)}]{quantum-simu-rev-sci}%
  \BibitemOpen
  \bibfield  {author} {\bibinfo {author} {\bibfnamefont {C.}~\bibnamefont
  {Gross}}\ and\ \bibinfo {author} {\bibfnamefont {I.}~\bibnamefont {Bloch}},\
  }\href {\doibase 10.1126/science.aal3837} {\bibfield  {journal} {\bibinfo
  {journal} {Science}\ }\textbf {\bibinfo {volume} {357}},\ \bibinfo {pages}
  {995} (\bibinfo {year} {2017})}\BibitemShut {NoStop}%
\bibitem [{\citenamefont {Dalibard}\ \emph {et~al.}(2011)\citenamefont
  {Dalibard}, \citenamefont {Gerbier}, \citenamefont {Juzeli{\=u}nas},\ and\
  \citenamefont {\"Ohberg}}]{laser-hop-rev-1}%
  \BibitemOpen
  \bibfield  {author} {\bibinfo {author} {\bibfnamefont {J.}~\bibnamefont
  {Dalibard}}, \bibinfo {author} {\bibfnamefont {F.}~\bibnamefont {Gerbier}},
  \bibinfo {author} {\bibfnamefont {G.}~\bibnamefont {Juzeli{\=u}nas}}, \ and\
  \bibinfo {author} {\bibfnamefont {P.}~\bibnamefont {\"Ohberg}},\ }\href
  {\doibase 10.1103/RevModPhys.83.1523} {\bibfield  {journal} {\bibinfo
  {journal} {Rev. Mod. Phys.}\ }\textbf {\bibinfo {volume} {83}},\ \bibinfo
  {pages} {1523} (\bibinfo {year} {2011})}\BibitemShut {NoStop}%
\bibitem [{\citenamefont {Goldman}\ \emph {et~al.}(2014)\citenamefont
  {Goldman}, \citenamefont {Juzeli{\=u}nas}, \citenamefont {{\"O}hberg},\ and\
  \citenamefont {Spielman}}]{laser-hop-rev-2}%
  \BibitemOpen
  \bibfield  {author} {\bibinfo {author} {\bibfnamefont {N.}~\bibnamefont
  {Goldman}}, \bibinfo {author} {\bibfnamefont {G.}~\bibnamefont
  {Juzeli{\=u}nas}}, \bibinfo {author} {\bibfnamefont {P.}~\bibnamefont
  {{\"O}hberg}}, \ and\ \bibinfo {author} {\bibfnamefont {I.~B.}\ \bibnamefont
  {Spielman}},\ }\href {\doibase 10.1088/0034-4885/77/12/126401} {\bibfield
  {journal} {\bibinfo  {journal} {Rep. Prog. Phys.}\ }\textbf {\bibinfo
  {volume} {77}},\ \bibinfo {pages} {126401} (\bibinfo {year}
  {2014})}\BibitemShut {NoStop}%
\bibitem [{\citenamefont {K\"ohler}\ \emph {et~al.}(2006)\citenamefont
  {K\"ohler}, \citenamefont {G\'oral},\ and\ \citenamefont
  {Julienne}}]{feshbach-rev-1}%
  \BibitemOpen
  \bibfield  {author} {\bibinfo {author} {\bibfnamefont {T.}~\bibnamefont
  {K\"ohler}}, \bibinfo {author} {\bibfnamefont {K.}~\bibnamefont {G\'oral}}, \
  and\ \bibinfo {author} {\bibfnamefont {P.~S.}\ \bibnamefont {Julienne}},\
  }\href {\doibase 10.1103/RevModPhys.78.1311} {\bibfield  {journal} {\bibinfo
  {journal} {Rev. Mod. Phys.}\ }\textbf {\bibinfo {volume} {78}},\ \bibinfo
  {pages} {1311} (\bibinfo {year} {2006})}\BibitemShut {NoStop}%
\bibitem [{\citenamefont {Chin}\ \emph {et~al.}(2010)\citenamefont {Chin},
  \citenamefont {Grimm}, \citenamefont {Julienne},\ and\ \citenamefont
  {Tiesinga}}]{feshbach-rev-2}%
  \BibitemOpen
  \bibfield  {author} {\bibinfo {author} {\bibfnamefont {C.}~\bibnamefont
  {Chin}}, \bibinfo {author} {\bibfnamefont {R.}~\bibnamefont {Grimm}},
  \bibinfo {author} {\bibfnamefont {P.}~\bibnamefont {Julienne}}, \ and\
  \bibinfo {author} {\bibfnamefont {E.}~\bibnamefont {Tiesinga}},\ }\href
  {\doibase 10.1103/RevModPhys.82.1225} {\bibfield  {journal} {\bibinfo
  {journal} {Rev. Mod. Phys.}\ }\textbf {\bibinfo {volume} {82}},\ \bibinfo
  {pages} {1225} (\bibinfo {year} {2010})}\BibitemShut {NoStop}%
\bibitem [{\citenamefont {Giorgini}\ \emph {et~al.}(2008)\citenamefont
  {Giorgini}, \citenamefont {Pitaevskii},\ and\ \citenamefont
  {Stringari}}]{fermi-gas-rev}%
  \BibitemOpen
  \bibfield  {author} {\bibinfo {author} {\bibfnamefont {S.}~\bibnamefont
  {Giorgini}}, \bibinfo {author} {\bibfnamefont {L.~P.}\ \bibnamefont
  {Pitaevskii}}, \ and\ \bibinfo {author} {\bibfnamefont {S.}~\bibnamefont
  {Stringari}},\ }\href {\doibase 10.1103/RevModPhys.80.1215} {\bibfield
  {journal} {\bibinfo  {journal} {Rev. Mod. Phys.}\ }\textbf {\bibinfo {volume}
  {80}},\ \bibinfo {pages} {1215} (\bibinfo {year} {2008})}\BibitemShut
  {NoStop}%
\bibitem [{\citenamefont {Ketterle}\ and\ \citenamefont
  {Zwierlein}(2008)}]{ketterle2008proceedings}%
  \BibitemOpen
  \bibfield  {author} {\bibinfo {author} {\bibfnamefont {W.}~\bibnamefont
  {Ketterle}}\ and\ \bibinfo {author} {\bibfnamefont {M.}~\bibnamefont
  {Zwierlein}},\ }\href@noop {} {\bibfield  {journal} {\bibinfo  {journal} {the
  Proceedings of the International School of Physics ``Enrico Fermi", Course
  CLXIV, edited by M. Inguscio, W. Ketterle, and C. Salomon}\ } (\bibinfo
  {year} {IOS Press, Amsterdam, 2008})}\BibitemShut {NoStop}%
\bibitem [{\citenamefont {Cui}\ \emph {et~al.}(2017)\citenamefont {Cui},
  \citenamefont {Shen}, \citenamefont {Deng}, \citenamefont {Dong},
  \citenamefont {Chen}, \citenamefont {L\"u}, \citenamefont {Gao},
  \citenamefont {Tey},\ and\ \citenamefont {You}}]{cold-atom-exp-d-wave}%
  \BibitemOpen
  \bibfield  {author} {\bibinfo {author} {\bibfnamefont {Y.}~\bibnamefont
  {Cui}}, \bibinfo {author} {\bibfnamefont {C.}~\bibnamefont {Shen}}, \bibinfo
  {author} {\bibfnamefont {M.}~\bibnamefont {Deng}}, \bibinfo {author}
  {\bibfnamefont {S.}~\bibnamefont {Dong}}, \bibinfo {author} {\bibfnamefont
  {C.}~\bibnamefont {Chen}}, \bibinfo {author} {\bibfnamefont {R.}~\bibnamefont
  {L\"u}}, \bibinfo {author} {\bibfnamefont {B.}~\bibnamefont {Gao}}, \bibinfo
  {author} {\bibfnamefont {M.~K.}\ \bibnamefont {Tey}}, \ and\ \bibinfo
  {author} {\bibfnamefont {L.}~\bibnamefont {You}},\ }\href {\doibase
  10.1103/PhysRevLett.119.203402} {\bibfield  {journal} {\bibinfo  {journal}
  {Phys. Rev. Lett.}\ }\textbf {\bibinfo {volume} {119}},\ \bibinfo {pages}
  {203402} (\bibinfo {year} {2017})}\BibitemShut {NoStop}%
\bibitem [{\citenamefont {Zhang}\ \emph {et~al.}(2015)\citenamefont {Zhang},
  \citenamefont {Lang},\ and\ \citenamefont {Zhou}}]{qi-d-wave}%
  \BibitemOpen
  \bibfield  {author} {\bibinfo {author} {\bibfnamefont {S.-L.}\ \bibnamefont
  {Zhang}}, \bibinfo {author} {\bibfnamefont {L.-J.}\ \bibnamefont {Lang}}, \
  and\ \bibinfo {author} {\bibfnamefont {Q.}~\bibnamefont {Zhou}},\ }\href
  {\doibase 10.1103/PhysRevLett.115.225301} {\bibfield  {journal} {\bibinfo
  {journal} {Phys. Rev. Lett.}\ }\textbf {\bibinfo {volume} {115}},\ \bibinfo
  {pages} {225301} (\bibinfo {year} {2015})}\BibitemShut {NoStop}%
\bibitem [{\citenamefont {Hao}\ \emph {et~al.}(2017)\citenamefont {Hao},
  \citenamefont {Guo},\ and\ \citenamefont {Zhang}}]{njp-d-wave}%
  \BibitemOpen
  \bibfield  {author} {\bibinfo {author} {\bibfnamefont {N.}~\bibnamefont
  {Hao}}, \bibinfo {author} {\bibfnamefont {H.}~\bibnamefont {Guo}}, \ and\
  \bibinfo {author} {\bibfnamefont {P.}~\bibnamefont {Zhang}},\ }\href
  {http://stacks.iop.org/1367-2630/19/i=8/a=083020} {\bibfield  {journal}
  {\bibinfo  {journal} {New J. Phys.}\ }\textbf {\bibinfo {volume} {19}},\
  \bibinfo {pages} {083020} (\bibinfo {year} {2017})}\BibitemShut {NoStop}%
\bibitem [{\citenamefont {Pistolesi}\ and\ \citenamefont
  {Nozi\`eres}(2002)}]{d-wave-NNN-1}%
  \BibitemOpen
  \bibfield  {author} {\bibinfo {author} {\bibfnamefont {F.}~\bibnamefont
  {Pistolesi}}\ and\ \bibinfo {author} {\bibfnamefont {P.}~\bibnamefont
  {Nozi\`eres}},\ }\href {\doibase 10.1103/PhysRevB.66.054501} {\bibfield
  {journal} {\bibinfo  {journal} {Phys. Rev. B}\ }\textbf {\bibinfo {volume}
  {66}},\ \bibinfo {pages} {054501} (\bibinfo {year} {2002})}\BibitemShut
  {NoStop}%
\bibitem [{\citenamefont {Gadsb\o{}lle}\ \emph {et~al.}(2012)\citenamefont
  {Gadsb\o{}lle}, \citenamefont {Song},\ and\ \citenamefont
  {Le~Hur}}]{d-wave-NNN-2}%
  \BibitemOpen
  \bibfield  {author} {\bibinfo {author} {\bibfnamefont {A.-L.}\ \bibnamefont
  {Gadsb\o{}lle}}, \bibinfo {author} {\bibfnamefont {H.~F.}\ \bibnamefont
  {Song}}, \ and\ \bibinfo {author} {\bibfnamefont {K.}~\bibnamefont
  {Le~Hur}},\ }\href {\doibase 10.1103/PhysRevA.85.051603} {\bibfield
  {journal} {\bibinfo  {journal} {Phys. Rev. A}\ }\textbf {\bibinfo {volume}
  {85}},\ \bibinfo {pages} {051603(R)} (\bibinfo {year} {2012})}\BibitemShut
  {NoStop}%
\bibitem [{\citenamefont {Rahav}\ \emph {et~al.}(2003)\citenamefont {Rahav},
  \citenamefont {Gilary},\ and\ \citenamefont {Fishman}}]{driven-theo-1}%
  \BibitemOpen
  \bibfield  {author} {\bibinfo {author} {\bibfnamefont {S.}~\bibnamefont
  {Rahav}}, \bibinfo {author} {\bibfnamefont {I.}~\bibnamefont {Gilary}}, \
  and\ \bibinfo {author} {\bibfnamefont {S.}~\bibnamefont {Fishman}},\ }\href
  {\doibase 10.1103/PhysRevA.68.013820} {\bibfield  {journal} {\bibinfo
  {journal} {Phys. Rev. A}\ }\textbf {\bibinfo {volume} {68}},\ \bibinfo
  {pages} {013820} (\bibinfo {year} {2003})}\BibitemShut {NoStop}%
\bibitem [{\citenamefont {Goldman}\ \emph {et~al.}(2015)\citenamefont
  {Goldman}, \citenamefont {Dalibard}, \citenamefont {Aidelsburger},\ and\
  \citenamefont {Cooper}}]{driven-theo-2}%
  \BibitemOpen
  \bibfield  {author} {\bibinfo {author} {\bibfnamefont {N.}~\bibnamefont
  {Goldman}}, \bibinfo {author} {\bibfnamefont {J.}~\bibnamefont {Dalibard}},
  \bibinfo {author} {\bibfnamefont {M.}~\bibnamefont {Aidelsburger}}, \ and\
  \bibinfo {author} {\bibfnamefont {N.~R.}\ \bibnamefont {Cooper}},\ }\href
  {\doibase 10.1103/PhysRevA.91.033632} {\bibfield  {journal} {\bibinfo
  {journal} {Phys. Rev. A}\ }\textbf {\bibinfo {volume} {91}},\ \bibinfo
  {pages} {033632} (\bibinfo {year} {2015})}\BibitemShut {NoStop}%
\bibitem [{\citenamefont {Goldman}\ and\ \citenamefont
  {Dalibard}(2014)}]{driven-theo-3}%
  \BibitemOpen
  \bibfield  {author} {\bibinfo {author} {\bibfnamefont {N.}~\bibnamefont
  {Goldman}}\ and\ \bibinfo {author} {\bibfnamefont {J.}~\bibnamefont
  {Dalibard}},\ }\href {\doibase 10.1103/PhysRevX.4.031027} {\bibfield
  {journal} {\bibinfo  {journal} {Phys. Rev. X}\ }\textbf {\bibinfo {volume}
  {4}},\ \bibinfo {pages} {031027} (\bibinfo {year} {2014})}\BibitemShut
  {NoStop}%
\bibitem [{\citenamefont {Eckardt}(2017)}]{floquet-rev}%
  \BibitemOpen
  \bibfield  {author} {\bibinfo {author} {\bibfnamefont {A.}~\bibnamefont
  {Eckardt}},\ }\href {\doibase 10.1103/RevModPhys.89.011004} {\bibfield
  {journal} {\bibinfo  {journal} {Rev. Mod. Phys.}\ }\textbf {\bibinfo {volume}
  {89}},\ \bibinfo {pages} {011004} (\bibinfo {year} {2017})}\BibitemShut
  {NoStop}%
\bibitem [{\citenamefont {Eckardt}\ \emph {et~al.}(2005)\citenamefont
  {Eckardt}, \citenamefont {Weiss},\ and\ \citenamefont
  {Holthaus}}]{driven-Eckardt-2005}%
  \BibitemOpen
  \bibfield  {author} {\bibinfo {author} {\bibfnamefont {A.}~\bibnamefont
  {Eckardt}}, \bibinfo {author} {\bibfnamefont {C.}~\bibnamefont {Weiss}}, \
  and\ \bibinfo {author} {\bibfnamefont {M.}~\bibnamefont {Holthaus}},\ }\href
  {\doibase 10.1103/PhysRevLett.95.260404} {\bibfield  {journal} {\bibinfo
  {journal} {Phys. Rev. Lett.}\ }\textbf {\bibinfo {volume} {95}},\ \bibinfo
  {pages} {260404} (\bibinfo {year} {2005})}\BibitemShut {NoStop}%
\bibitem [{\citenamefont {Lignier}\ \emph {et~al.}(2007)\citenamefont
  {Lignier}, \citenamefont {Sias}, \citenamefont {Ciampini}, \citenamefont
  {Singh}, \citenamefont {Zenesini}, \citenamefont {Morsch},\ and\
  \citenamefont {Arimondo}}]{driven-Lignier-2007}%
  \BibitemOpen
  \bibfield  {author} {\bibinfo {author} {\bibfnamefont {H.}~\bibnamefont
  {Lignier}}, \bibinfo {author} {\bibfnamefont {C.}~\bibnamefont {Sias}},
  \bibinfo {author} {\bibfnamefont {D.}~\bibnamefont {Ciampini}}, \bibinfo
  {author} {\bibfnamefont {Y.}~\bibnamefont {Singh}}, \bibinfo {author}
  {\bibfnamefont {A.}~\bibnamefont {Zenesini}}, \bibinfo {author}
  {\bibfnamefont {O.}~\bibnamefont {Morsch}}, \ and\ \bibinfo {author}
  {\bibfnamefont {E.}~\bibnamefont {Arimondo}},\ }\href {\doibase
  10.1103/PhysRevLett.99.220403} {\bibfield  {journal} {\bibinfo  {journal}
  {Phys. Rev. Lett.}\ }\textbf {\bibinfo {volume} {99}},\ \bibinfo {pages}
  {220403} (\bibinfo {year} {2007})}\BibitemShut {NoStop}%
\bibitem [{\citenamefont {Lindner}\ \emph {et~al.}(2011)\citenamefont
  {Lindner}, \citenamefont {Refael},\ and\ \citenamefont
  {Galitski}}]{lindner2011floquet}%
  \BibitemOpen
  \bibfield  {author} {\bibinfo {author} {\bibfnamefont {N.~H.}\ \bibnamefont
  {Lindner}}, \bibinfo {author} {\bibfnamefont {G.}~\bibnamefont {Refael}}, \
  and\ \bibinfo {author} {\bibfnamefont {V.}~\bibnamefont {Galitski}},\ }\href
  {\doibase doi:10.1038/nphys1926} {\bibfield  {journal} {\bibinfo  {journal}
  {Nature Physics}\ }\textbf {\bibinfo {volume} {7}},\ \bibinfo {pages} {490}
  (\bibinfo {year} {2011})}\BibitemShut {NoStop}%
\bibitem [{\citenamefont {Liu}\ \emph {et~al.}(2012)\citenamefont {Liu},
  \citenamefont {Hao}, \citenamefont {Zhu},\ and\ \citenamefont
  {Liu}}]{driven-Liu-2012}%
  \BibitemOpen
  \bibfield  {author} {\bibinfo {author} {\bibfnamefont {G.}~\bibnamefont
  {Liu}}, \bibinfo {author} {\bibfnamefont {N.}~\bibnamefont {Hao}}, \bibinfo
  {author} {\bibfnamefont {S.-L.}\ \bibnamefont {Zhu}}, \ and\ \bibinfo
  {author} {\bibfnamefont {W.~M.}\ \bibnamefont {Liu}},\ }\href {\doibase
  10.1103/PhysRevA.86.013639} {\bibfield  {journal} {\bibinfo  {journal} {Phys.
  Rev. A}\ }\textbf {\bibinfo {volume} {86}},\ \bibinfo {pages} {013639}
  (\bibinfo {year} {2012})}\BibitemShut {NoStop}%
\bibitem [{\citenamefont {Zheng}\ and\ \citenamefont
  {Zhai}(2014)}]{driven-Zheng-2014}%
  \BibitemOpen
  \bibfield  {author} {\bibinfo {author} {\bibfnamefont {W.}~\bibnamefont
  {Zheng}}\ and\ \bibinfo {author} {\bibfnamefont {H.}~\bibnamefont {Zhai}},\
  }\href {\doibase 10.1103/PhysRevA.89.061603} {\bibfield  {journal} {\bibinfo
  {journal} {Phys. Rev. A}\ }\textbf {\bibinfo {volume} {89}},\ \bibinfo
  {pages} {061603(R)} (\bibinfo {year} {2014})}\BibitemShut {NoStop}%
\bibitem [{\citenamefont {Struck}\ \emph {et~al.}(2014)\citenamefont {Struck},
  \citenamefont {Simonet},\ and\ \citenamefont
  {Sengstock}}]{driven-Struck-2014}%
  \BibitemOpen
  \bibfield  {author} {\bibinfo {author} {\bibfnamefont {J.}~\bibnamefont
  {Struck}}, \bibinfo {author} {\bibfnamefont {J.}~\bibnamefont {Simonet}}, \
  and\ \bibinfo {author} {\bibfnamefont {K.}~\bibnamefont {Sengstock}},\ }\href
  {\doibase 10.1103/PhysRevA.90.031601} {\bibfield  {journal} {\bibinfo
  {journal} {Phys. Rev. A}\ }\textbf {\bibinfo {volume} {90}},\ \bibinfo
  {pages} {031601(R)} (\bibinfo {year} {2014})}\BibitemShut {NoStop}%
\bibitem [{\citenamefont {Zheng}\ \emph {et~al.}(2015)\citenamefont {Zheng},
  \citenamefont {Qu}, \citenamefont {Zou},\ and\ \citenamefont
  {Zhang}}]{driven-Zheng-2015}%
  \BibitemOpen
  \bibfield  {author} {\bibinfo {author} {\bibfnamefont {Z.}~\bibnamefont
  {Zheng}}, \bibinfo {author} {\bibfnamefont {C.}~\bibnamefont {Qu}}, \bibinfo
  {author} {\bibfnamefont {X.}~\bibnamefont {Zou}}, \ and\ \bibinfo {author}
  {\bibfnamefont {C.}~\bibnamefont {Zhang}},\ }\href {\doibase
  10.1103/PhysRevA.91.063626} {\bibfield  {journal} {\bibinfo  {journal} {Phys.
  Rev. A}\ }\textbf {\bibinfo {volume} {91}},\ \bibinfo {pages} {063626}
  (\bibinfo {year} {2015})}\BibitemShut {NoStop}%
\bibitem [{\citenamefont {Zheng}\ \emph {et~al.}(2016)\citenamefont {Zheng},
  \citenamefont {Qu}, \citenamefont {Zou},\ and\ \citenamefont
  {Zhang}}]{driven-Zheng-2016}%
  \BibitemOpen
  \bibfield  {author} {\bibinfo {author} {\bibfnamefont {Z.}~\bibnamefont
  {Zheng}}, \bibinfo {author} {\bibfnamefont {C.}~\bibnamefont {Qu}}, \bibinfo
  {author} {\bibfnamefont {X.}~\bibnamefont {Zou}}, \ and\ \bibinfo {author}
  {\bibfnamefont {C.}~\bibnamefont {Zhang}},\ }\href {\doibase
  10.1103/PhysRevLett.116.120403} {\bibfield  {journal} {\bibinfo  {journal}
  {Phys. Rev. Lett.}\ }\textbf {\bibinfo {volume} {116}},\ \bibinfo {pages}
  {120403} (\bibinfo {year} {2016})}\BibitemShut {NoStop}%
\bibitem [{\citenamefont {Meinert}\ \emph {et~al.}(2016)\citenamefont
  {Meinert}, \citenamefont {Mark}, \citenamefont {Lauber}, \citenamefont
  {Daley},\ and\ \citenamefont {N\"agerl}}]{driven-Meinert-2016}%
  \BibitemOpen
  \bibfield  {author} {\bibinfo {author} {\bibfnamefont {F.}~\bibnamefont
  {Meinert}}, \bibinfo {author} {\bibfnamefont {M.~J.}\ \bibnamefont {Mark}},
  \bibinfo {author} {\bibfnamefont {K.}~\bibnamefont {Lauber}}, \bibinfo
  {author} {\bibfnamefont {A.~J.}\ \bibnamefont {Daley}}, \ and\ \bibinfo
  {author} {\bibfnamefont {H.-C.}\ \bibnamefont {N\"agerl}},\ }\href {\doibase
  10.1103/PhysRevLett.116.205301} {\bibfield  {journal} {\bibinfo  {journal}
  {Phys. Rev. Lett.}\ }\textbf {\bibinfo {volume} {116}},\ \bibinfo {pages}
  {205301} (\bibinfo {year} {2016})}\BibitemShut {NoStop}%
\bibitem [{\citenamefont {Nocera}\ \emph {et~al.}(2017)\citenamefont {Nocera},
  \citenamefont {Polkovnikov},\ and\ \citenamefont
  {Feiguin}}]{driven-Nocera-2017}%
  \BibitemOpen
  \bibfield  {author} {\bibinfo {author} {\bibfnamefont {A.}~\bibnamefont
  {Nocera}}, \bibinfo {author} {\bibfnamefont {A.}~\bibnamefont {Polkovnikov}},
  \ and\ \bibinfo {author} {\bibfnamefont {A.~E.}\ \bibnamefont {Feiguin}},\
  }\href {\doibase 10.1103/PhysRevA.95.023601} {\bibfield  {journal} {\bibinfo
  {journal} {Phys. Rev. A}\ }\textbf {\bibinfo {volume} {95}},\ \bibinfo
  {pages} {023601} (\bibinfo {year} {2017})}\BibitemShut {NoStop}%
\bibitem [{\citenamefont {Xu}\ \emph {et~al.}(2017)\citenamefont {Xu},
  \citenamefont {Zhang},\ and\ \citenamefont {Chen}}]{driven-Xu-2017}%
  \BibitemOpen
  \bibfield  {author} {\bibinfo {author} {\bibfnamefont {Z.}~\bibnamefont
  {Xu}}, \bibinfo {author} {\bibfnamefont {Y.}~\bibnamefont {Zhang}}, \ and\
  \bibinfo {author} {\bibfnamefont {S.}~\bibnamefont {Chen}},\ }\href {\doibase
  10.1103/PhysRevA.96.013606} {\bibfield  {journal} {\bibinfo  {journal} {Phys.
  Rev. A}\ }\textbf {\bibinfo {volume} {96}},\ \bibinfo {pages} {013606}
  (\bibinfo {year} {2017})}\BibitemShut {NoStop}%
\bibitem [{\citenamefont {Gra\ss{}}\ \emph {et~al.}(2018)\citenamefont
  {Gra\ss{}}, \citenamefont {Celi}, \citenamefont {Pagano},\ and\ \citenamefont
  {Lewenstein}}]{driven-Grass-2018}%
  \BibitemOpen
  \bibfield  {author} {\bibinfo {author} {\bibfnamefont {T.}~\bibnamefont
  {Gra\ss{}}}, \bibinfo {author} {\bibfnamefont {A.}~\bibnamefont {Celi}},
  \bibinfo {author} {\bibfnamefont {G.}~\bibnamefont {Pagano}}, \ and\ \bibinfo
  {author} {\bibfnamefont {M.}~\bibnamefont {Lewenstein}},\ }\href {\doibase
  10.1103/PhysRevA.97.010302} {\bibfield  {journal} {\bibinfo  {journal} {Phys.
  Rev. A}\ }\textbf {\bibinfo {volume} {97}},\ \bibinfo {pages} {010302(R)}
  (\bibinfo {year} {2018})}\BibitemShut {NoStop}%
\bibitem [{\citenamefont {Zhou}\ \emph {et~al.}(2018)\citenamefont {Zhou},
  \citenamefont {Wu}, \citenamefont {Guo}, \citenamefont {Wang}, \citenamefont
  {Pu},\ and\ \citenamefont {Zhou}}]{driven-Zhou-2018}%
  \BibitemOpen
  \bibfield  {author} {\bibinfo {author} {\bibfnamefont {X.-F.}\ \bibnamefont
  {Zhou}}, \bibinfo {author} {\bibfnamefont {C.}~\bibnamefont {Wu}}, \bibinfo
  {author} {\bibfnamefont {G.-C.}\ \bibnamefont {Guo}}, \bibinfo {author}
  {\bibfnamefont {R.}~\bibnamefont {Wang}}, \bibinfo {author} {\bibfnamefont
  {H.}~\bibnamefont {Pu}}, \ and\ \bibinfo {author} {\bibfnamefont {Z.-W.}\
  \bibnamefont {Zhou}},\ }\href {\doibase 10.1103/PhysRevLett.120.130402}
  {\bibfield  {journal} {\bibinfo  {journal} {Phys. Rev. Lett.}\ }\textbf
  {\bibinfo {volume} {120}},\ \bibinfo {pages} {130402} (\bibinfo {year}
  {2018})}\BibitemShut {NoStop}%
\bibitem [{\citenamefont {Messer}\ \emph {et~al.}(2018)\citenamefont {Messer},
  \citenamefont {Sandholzer}, \citenamefont {G\"org}, \citenamefont {Minguzzi},
  \citenamefont {Desbuquois},\ and\ \citenamefont
  {Esslinger}}]{driven-Messer-2018}%
  \BibitemOpen
  \bibfield  {author} {\bibinfo {author} {\bibfnamefont {M.}~\bibnamefont
  {Messer}}, \bibinfo {author} {\bibfnamefont {K.}~\bibnamefont {Sandholzer}},
  \bibinfo {author} {\bibfnamefont {F.}~\bibnamefont {G\"org}}, \bibinfo
  {author} {\bibfnamefont {J.}~\bibnamefont {Minguzzi}}, \bibinfo {author}
  {\bibfnamefont {R.}~\bibnamefont {Desbuquois}}, \ and\ \bibinfo {author}
  {\bibfnamefont {T.}~\bibnamefont {Esslinger}},\ }\href {\doibase
  10.1103/PhysRevLett.121.233603} {\bibfield  {journal} {\bibinfo  {journal}
  {Phys. Rev. Lett.}\ }\textbf {\bibinfo {volume} {121}},\ \bibinfo {pages}
  {233603} (\bibinfo {year} {2018})}\BibitemShut {NoStop}%
\bibitem [{\citenamefont {Zheng}\ and\ \citenamefont
  {Wang}(2019)}]{Zheng_2019}%
  \BibitemOpen
  \bibfield  {author} {\bibinfo {author} {\bibfnamefont {Z.}~\bibnamefont
  {Zheng}}\ and\ \bibinfo {author} {\bibfnamefont {Z.~D.}\ \bibnamefont
  {Wang}},\ }\href {\doibase 10.1088/1361-6455/ab08df} {\bibfield  {journal}
  {\bibinfo  {journal} {J. Phys. B: At. Mol. Opt. Phys.}\ }\textbf {\bibinfo
  {volume} {52}},\ \bibinfo {pages} {075301} (\bibinfo {year}
  {2019})}\BibitemShut {NoStop}%
\bibitem [{\citenamefont {Itin}\ and\ \citenamefont
  {Katsnelson}(2015)}]{driven-cor-hop-1}%
  \BibitemOpen
  \bibfield  {author} {\bibinfo {author} {\bibfnamefont {A.~P.}\ \bibnamefont
  {Itin}}\ and\ \bibinfo {author} {\bibfnamefont {M.~I.}\ \bibnamefont
  {Katsnelson}},\ }\href {\doibase 10.1103/PhysRevLett.115.075301} {\bibfield
  {journal} {\bibinfo  {journal} {Phys. Rev. Lett.}\ }\textbf {\bibinfo
  {volume} {115}},\ \bibinfo {pages} {075301} (\bibinfo {year}
  {2015})}\BibitemShut {NoStop}%
\bibitem [{\citenamefont {Chen}\ \emph {et~al.}(2011)\citenamefont {Chen},
  \citenamefont {Nascimb\`ene}, \citenamefont {Aidelsburger}, \citenamefont
  {Atala}, \citenamefont {Trotzky},\ and\ \citenamefont
  {Bloch}}]{driven-Chen-2011}%
  \BibitemOpen
  \bibfield  {author} {\bibinfo {author} {\bibfnamefont {Y.-A.}\ \bibnamefont
  {Chen}}, \bibinfo {author} {\bibfnamefont {S.}~\bibnamefont {Nascimb\`ene}},
  \bibinfo {author} {\bibfnamefont {M.}~\bibnamefont {Aidelsburger}}, \bibinfo
  {author} {\bibfnamefont {M.}~\bibnamefont {Atala}}, \bibinfo {author}
  {\bibfnamefont {S.}~\bibnamefont {Trotzky}}, \ and\ \bibinfo {author}
  {\bibfnamefont {I.}~\bibnamefont {Bloch}},\ }\href {\doibase
  10.1103/PhysRevLett.107.210405} {\bibfield  {journal} {\bibinfo  {journal}
  {Phys. Rev. Lett.}\ }\textbf {\bibinfo {volume} {107}},\ \bibinfo {pages}
  {210405} (\bibinfo {year} {2011})}\BibitemShut {NoStop}%
\bibitem [{\citenamefont {Bukov}\ \emph {et~al.}(2016)\citenamefont {Bukov},
  \citenamefont {Kolodrubetz},\ and\ \citenamefont
  {Polkovnikov}}]{driven-Bukov-2016}%
  \BibitemOpen
  \bibfield  {author} {\bibinfo {author} {\bibfnamefont {M.}~\bibnamefont
  {Bukov}}, \bibinfo {author} {\bibfnamefont {M.}~\bibnamefont {Kolodrubetz}},
  \ and\ \bibinfo {author} {\bibfnamefont {A.}~\bibnamefont {Polkovnikov}},\
  }\href {\doibase 10.1103/PhysRevLett.116.125301} {\bibfield  {journal}
  {\bibinfo  {journal} {Phys. Rev. Lett.}\ }\textbf {\bibinfo {volume} {116}},\
  \bibinfo {pages} {125301} (\bibinfo {year} {2016})}\BibitemShut {NoStop}%
\bibitem [{\citenamefont {Zhang}\ and\ \citenamefont
  {Zhou}(2017)}]{driven-Zhang-2017}%
  \BibitemOpen
  \bibfield  {author} {\bibinfo {author} {\bibfnamefont {S.-L.}\ \bibnamefont
  {Zhang}}\ and\ \bibinfo {author} {\bibfnamefont {Q.}~\bibnamefont {Zhou}},\
  }\href {\doibase 10.1088/1361-6455/aa8c5a} {\bibfield  {journal} {\bibinfo
  {journal} {J. Phys. B: At. Mol. Opt. Phys.}\ }\textbf {\bibinfo {volume}
  {50}},\ \bibinfo {pages} {222001} (\bibinfo {year} {2017})}\BibitemShut
  {NoStop}%
\bibitem [{\citenamefont {Wang}\ \emph
  {et~al.}(2018{\natexlab{b}})\citenamefont {Wang}, \citenamefont {\"Unal},\
  and\ \citenamefont {Eckardt}}]{driven-Botao-2018}%
  \BibitemOpen
  \bibfield  {author} {\bibinfo {author} {\bibfnamefont {B.}~\bibnamefont
  {Wang}}, \bibinfo {author} {\bibfnamefont {F.~N.}\ \bibnamefont {\"Unal}}, \
  and\ \bibinfo {author} {\bibfnamefont {A.}~\bibnamefont {Eckardt}},\ }\href
  {\doibase 10.1103/PhysRevLett.120.243602} {\bibfield  {journal} {\bibinfo
  {journal} {Phys. Rev. Lett.}\ }\textbf {\bibinfo {volume} {120}},\ \bibinfo
  {pages} {243602} (\bibinfo {year} {2018}{\natexlab{b}})}\BibitemShut
  {NoStop}%
\bibitem [{\citenamefont {Dickerscheid}\ \emph {et~al.}(2005)\citenamefont
  {Dickerscheid}, \citenamefont {Al~Khawaja}, \citenamefont {van Oosten},\ and\
  \citenamefont {Stoof}}]{lattice-feshbach-1}%
  \BibitemOpen
  \bibfield  {author} {\bibinfo {author} {\bibfnamefont {D.~B.~M.}\
  \bibnamefont {Dickerscheid}}, \bibinfo {author} {\bibfnamefont
  {U.}~\bibnamefont {Al~Khawaja}}, \bibinfo {author} {\bibfnamefont
  {D.}~\bibnamefont {van Oosten}}, \ and\ \bibinfo {author} {\bibfnamefont
  {H.~T.~C.}\ \bibnamefont {Stoof}},\ }\href {\doibase
  10.1103/PhysRevA.71.043604} {\bibfield  {journal} {\bibinfo  {journal} {Phys.
  Rev. A}\ }\textbf {\bibinfo {volume} {71}},\ \bibinfo {pages} {043604}
  (\bibinfo {year} {2005})}\BibitemShut {NoStop}%
\bibitem [{\citenamefont {Duan}(2005)}]{lattice-feshbach-2}%
  \BibitemOpen
  \bibfield  {author} {\bibinfo {author} {\bibfnamefont {L.-M.}\ \bibnamefont
  {Duan}},\ }\href {\doibase 10.1103/PhysRevLett.95.243202} {\bibfield
  {journal} {\bibinfo  {journal} {Phys. Rev. Lett.}\ }\textbf {\bibinfo
  {volume} {95}},\ \bibinfo {pages} {243202} (\bibinfo {year}
  {2005})}\BibitemShut {NoStop}%
\bibitem [{\citenamefont {Shen}\ \emph {et~al.}(2012)\citenamefont {Shen},
  \citenamefont {Radzihovsky},\ and\ \citenamefont
  {Gurarie}}]{lattice-crossover}%
  \BibitemOpen
  \bibfield  {author} {\bibinfo {author} {\bibfnamefont {Z.}~\bibnamefont
  {Shen}}, \bibinfo {author} {\bibfnamefont {L.}~\bibnamefont {Radzihovsky}}, \
  and\ \bibinfo {author} {\bibfnamefont {V.}~\bibnamefont {Gurarie}},\ }\href
  {\doibase 10.1103/PhysRevLett.109.245302} {\bibfield  {journal} {\bibinfo
  {journal} {Phys. Rev. Lett.}\ }\textbf {\bibinfo {volume} {109}},\ \bibinfo
  {pages} {245302} (\bibinfo {year} {2012})}\BibitemShut {NoStop}%
\bibitem [{\citenamefont {Botelho}\ and\ \citenamefont {S\'a~de
  Melo}(2006)}]{bkt1}%
  \BibitemOpen
  \bibfield  {author} {\bibinfo {author} {\bibfnamefont {S.~S.}\ \bibnamefont
  {Botelho}}\ and\ \bibinfo {author} {\bibfnamefont {C.~A.~R.}\ \bibnamefont
  {S\'a~de Melo}},\ }\href {\doibase 10.1103/PhysRevLett.96.040404} {\bibfield
  {journal} {\bibinfo  {journal} {Phys. Rev. Lett.}\ }\textbf {\bibinfo
  {volume} {96}},\ \bibinfo {pages} {040404} (\bibinfo {year}
  {2006})}\BibitemShut {NoStop}%
\bibitem [{\citenamefont {He}\ and\ \citenamefont {Huang}(2012)}]{bkt2}%
  \BibitemOpen
  \bibfield  {author} {\bibinfo {author} {\bibfnamefont {L.}~\bibnamefont
  {He}}\ and\ \bibinfo {author} {\bibfnamefont {X.-G.}\ \bibnamefont {Huang}},\
  }\href {\doibase 10.1103/PhysRevLett.108.145302} {\bibfield  {journal}
  {\bibinfo  {journal} {Phys. Rev. Lett.}\ }\textbf {\bibinfo {volume} {108}},\
  \bibinfo {pages} {145302} (\bibinfo {year} {2012})}\BibitemShut {NoStop}%
\bibitem [{\citenamefont {Gong}\ \emph {et~al.}(2012)\citenamefont {Gong},
  \citenamefont {Chen}, \citenamefont {Jia},\ and\ \citenamefont
  {Zhang}}]{bkt3}%
  \BibitemOpen
  \bibfield  {author} {\bibinfo {author} {\bibfnamefont {M.}~\bibnamefont
  {Gong}}, \bibinfo {author} {\bibfnamefont {G.}~\bibnamefont {Chen}}, \bibinfo
  {author} {\bibfnamefont {S.}~\bibnamefont {Jia}}, \ and\ \bibinfo {author}
  {\bibfnamefont {C.}~\bibnamefont {Zhang}},\ }\href {\doibase
  10.1103/PhysRevLett.109.105302} {\bibfield  {journal} {\bibinfo  {journal}
  {Phys. Rev. Lett.}\ }\textbf {\bibinfo {volume} {109}},\ \bibinfo {pages}
  {105302} (\bibinfo {year} {2012})}\BibitemShut {NoStop}%
\bibitem [{\citenamefont {Xu}\ and\ \citenamefont {Zhang}(2015)}]{bkt4}%
  \BibitemOpen
  \bibfield  {author} {\bibinfo {author} {\bibfnamefont {Y.}~\bibnamefont
  {Xu}}\ and\ \bibinfo {author} {\bibfnamefont {C.}~\bibnamefont {Zhang}},\
  }\href {\doibase 10.1103/PhysRevLett.114.110401} {\bibfield  {journal}
  {\bibinfo  {journal} {Phys. Rev. Lett.}\ }\textbf {\bibinfo {volume} {114}},\
  \bibinfo {pages} {110401} (\bibinfo {year} {2015})}\BibitemShut {NoStop}%
\bibitem [{\citenamefont {Yin}\ \emph {et~al.}(2014)\citenamefont {Yin},
  \citenamefont {Martikainen},\ and\ \citenamefont {T\"orm\"a}}]{bkt5}%
  \BibitemOpen
  \bibfield  {author} {\bibinfo {author} {\bibfnamefont {S.}~\bibnamefont
  {Yin}}, \bibinfo {author} {\bibfnamefont {J.-P.}\ \bibnamefont
  {Martikainen}}, \ and\ \bibinfo {author} {\bibfnamefont {P.}~\bibnamefont
  {T\"orm\"a}},\ }\href {\doibase 10.1103/PhysRevB.89.014507} {\bibfield
  {journal} {\bibinfo  {journal} {Phys. Rev. B}\ }\textbf {\bibinfo {volume}
  {89}},\ \bibinfo {pages} {014507} (\bibinfo {year} {2014})}\BibitemShut
  {NoStop}%
\bibitem [{\citenamefont {Berezinskii}(1971)}]{bkt-origin-1}%
  \BibitemOpen
  \bibfield  {author} {\bibinfo {author} {\bibfnamefont {V.}~\bibnamefont
  {Berezinskii}},\ }\href@noop {} {\bibfield  {journal} {\bibinfo  {journal}
  {Sov. Phys. JETP}\ }\textbf {\bibinfo {volume} {32}},\ \bibinfo {pages} {493}
  (\bibinfo {year} {1971})}\BibitemShut {NoStop}%
\bibitem [{\citenamefont {Kosterlitz}\ and\ \citenamefont
  {Thouless}(1972)}]{bkt-origin-2}%
  \BibitemOpen
  \bibfield  {author} {\bibinfo {author} {\bibfnamefont {J.}~\bibnamefont
  {Kosterlitz}}\ and\ \bibinfo {author} {\bibfnamefont {D.}~\bibnamefont
  {Thouless}},\ }\href {\doibase 10.1088/0022-3719/5/11/002} {\bibfield
  {journal} {\bibinfo  {journal} {J. Phys. C}\ }\textbf {\bibinfo {volume}
  {5}},\ \bibinfo {pages} {L124} (\bibinfo {year} {1972})}\BibitemShut
  {NoStop}%
\bibitem [{\citenamefont {Kosterlitz}\ and\ \citenamefont
  {Thouless}(1973)}]{bkt-origin-3}%
  \BibitemOpen
  \bibfield  {author} {\bibinfo {author} {\bibfnamefont {J.~M.}\ \bibnamefont
  {Kosterlitz}}\ and\ \bibinfo {author} {\bibfnamefont {D.~J.}\ \bibnamefont
  {Thouless}},\ }\href {\doibase 10.1088/0022-3719/6/7/010} {\bibfield
  {journal} {\bibinfo  {journal} {J. Phys. C}\ }\textbf {\bibinfo {volume}
  {6}},\ \bibinfo {pages} {1181} (\bibinfo {year} {1973})}\BibitemShut
  {NoStop}%
\bibitem [{\citenamefont {Qi}\ \emph {et~al.}(2008)\citenamefont {Qi},
  \citenamefont {Hughes},\ and\ \citenamefont {Zhang}}]{2d-topo-ins}%
  \BibitemOpen
  \bibfield  {author} {\bibinfo {author} {\bibfnamefont {X.-L.}\ \bibnamefont
  {Qi}}, \bibinfo {author} {\bibfnamefont {T.~L.}\ \bibnamefont {Hughes}}, \
  and\ \bibinfo {author} {\bibfnamefont {S.-C.}\ \bibnamefont {Zhang}},\ }\href
  {\doibase 10.1103/PhysRevB.78.195424} {\bibfield  {journal} {\bibinfo
  {journal} {Phys. Rev. B}\ }\textbf {\bibinfo {volume} {78}},\ \bibinfo
  {pages} {195424} (\bibinfo {year} {2008})}\BibitemShut {NoStop}%
\bibitem [{\citenamefont {Wu}\ \emph {et~al.}(2016)\citenamefont {Wu},
  \citenamefont {Zhang}, \citenamefont {Sun}, \citenamefont {Xu}, \citenamefont
  {Wang}, \citenamefont {Ji}, \citenamefont {Deng}, \citenamefont {Chen},
  \citenamefont {Liu},\ and\ \citenamefont {Pan}}]{2d-soc-exp-1}%
  \BibitemOpen
  \bibfield  {author} {\bibinfo {author} {\bibfnamefont {Z.}~\bibnamefont
  {Wu}}, \bibinfo {author} {\bibfnamefont {L.}~\bibnamefont {Zhang}}, \bibinfo
  {author} {\bibfnamefont {W.}~\bibnamefont {Sun}}, \bibinfo {author}
  {\bibfnamefont {X.-T.}\ \bibnamefont {Xu}}, \bibinfo {author} {\bibfnamefont
  {B.-Z.}\ \bibnamefont {Wang}}, \bibinfo {author} {\bibfnamefont {S.-C.}\
  \bibnamefont {Ji}}, \bibinfo {author} {\bibfnamefont {Y.}~\bibnamefont
  {Deng}}, \bibinfo {author} {\bibfnamefont {S.}~\bibnamefont {Chen}}, \bibinfo
  {author} {\bibfnamefont {X.-J.}\ \bibnamefont {Liu}}, \ and\ \bibinfo
  {author} {\bibfnamefont {J.-W.}\ \bibnamefont {Pan}},\ }\href
  {http://doi.org/10.1126/science.aaf6689} {\bibfield  {journal} {\bibinfo
  {journal} {Science}\ }\textbf {\bibinfo {volume} {354}},\ \bibinfo {pages}
  {83} (\bibinfo {year} {2016})}\BibitemShut {NoStop}%
\bibitem [{\citenamefont {Meng}\ \emph {et~al.}(2016)\citenamefont {Meng},
  \citenamefont {Huang}, \citenamefont {Peng}, \citenamefont {Li},
  \citenamefont {Chen}, \citenamefont {Xu}, \citenamefont {Zhang},
  \citenamefont {Wang},\ and\ \citenamefont {Zhang}}]{2d-soc-exp-2}%
  \BibitemOpen
  \bibfield  {author} {\bibinfo {author} {\bibfnamefont {Z.}~\bibnamefont
  {Meng}}, \bibinfo {author} {\bibfnamefont {L.}~\bibnamefont {Huang}},
  \bibinfo {author} {\bibfnamefont {P.}~\bibnamefont {Peng}}, \bibinfo {author}
  {\bibfnamefont {D.}~\bibnamefont {Li}}, \bibinfo {author} {\bibfnamefont
  {L.}~\bibnamefont {Chen}}, \bibinfo {author} {\bibfnamefont {Y.}~\bibnamefont
  {Xu}}, \bibinfo {author} {\bibfnamefont {C.}~\bibnamefont {Zhang}}, \bibinfo
  {author} {\bibfnamefont {P.}~\bibnamefont {Wang}}, \ and\ \bibinfo {author}
  {\bibfnamefont {J.}~\bibnamefont {Zhang}},\ }\href {\doibase
  10.1103/PhysRevLett.117.235304} {\bibfield  {journal} {\bibinfo  {journal}
  {Phys. Rev. Lett.}\ }\textbf {\bibinfo {volume} {117}},\ \bibinfo {pages}
  {235304} (\bibinfo {year} {2016})}\BibitemShut {NoStop}%
\bibitem [{\citenamefont {Aidelsburger}\ \emph {et~al.}(2013)\citenamefont
  {Aidelsburger}, \citenamefont {Atala}, \citenamefont {Lohse}, \citenamefont
  {Barreiro}, \citenamefont {Paredes},\ and\ \citenamefont
  {Bloch}}]{laser-hop-1}%
  \BibitemOpen
  \bibfield  {author} {\bibinfo {author} {\bibfnamefont {M.}~\bibnamefont
  {Aidelsburger}}, \bibinfo {author} {\bibfnamefont {M.}~\bibnamefont {Atala}},
  \bibinfo {author} {\bibfnamefont {M.}~\bibnamefont {Lohse}}, \bibinfo
  {author} {\bibfnamefont {J.~T.}\ \bibnamefont {Barreiro}}, \bibinfo {author}
  {\bibfnamefont {B.}~\bibnamefont {Paredes}}, \ and\ \bibinfo {author}
  {\bibfnamefont {I.}~\bibnamefont {Bloch}},\ }\href {\doibase
  10.1103/PhysRevLett.111.185301} {\bibfield  {journal} {\bibinfo  {journal}
  {Phys. Rev. Lett.}\ }\textbf {\bibinfo {volume} {111}},\ \bibinfo {pages}
  {185301} (\bibinfo {year} {2013})}\BibitemShut {NoStop}%
\bibitem [{\citenamefont {Miyake}\ \emph {et~al.}(2013)\citenamefont {Miyake},
  \citenamefont {Siviloglou}, \citenamefont {Kennedy}, \citenamefont {Burton},\
  and\ \citenamefont {Ketterle}}]{laser-hop-2}%
  \BibitemOpen
  \bibfield  {author} {\bibinfo {author} {\bibfnamefont {H.}~\bibnamefont
  {Miyake}}, \bibinfo {author} {\bibfnamefont {G.~A.}\ \bibnamefont
  {Siviloglou}}, \bibinfo {author} {\bibfnamefont {C.~J.}\ \bibnamefont
  {Kennedy}}, \bibinfo {author} {\bibfnamefont {W.~C.}\ \bibnamefont {Burton}},
  \ and\ \bibinfo {author} {\bibfnamefont {W.}~\bibnamefont {Ketterle}},\
  }\href {\doibase 10.1103/PhysRevLett.111.185302} {\bibfield  {journal}
  {\bibinfo  {journal} {Phys. Rev. Lett.}\ }\textbf {\bibinfo {volume} {111}},\
  \bibinfo {pages} {185302} (\bibinfo {year} {2013})}\BibitemShut {NoStop}%
\bibitem [{\citenamefont {DeMarco}\ \emph {et~al.}(2001)\citenamefont
  {DeMarco}, \citenamefont {Papp},\ and\ \citenamefont {Jin}}]{pauli-blocking}%
  \BibitemOpen
  \bibfield  {author} {\bibinfo {author} {\bibfnamefont {B.}~\bibnamefont
  {DeMarco}}, \bibinfo {author} {\bibfnamefont {S.~B.}\ \bibnamefont {Papp}}, \
  and\ \bibinfo {author} {\bibfnamefont {D.~S.}\ \bibnamefont {Jin}},\ }\href
  {\doibase 10.1103/PhysRevLett.86.5409} {\bibfield  {journal} {\bibinfo
  {journal} {Phys. Rev. Lett.}\ }\textbf {\bibinfo {volume} {86}},\ \bibinfo
  {pages} {5409} (\bibinfo {year} {2001})}\BibitemShut {NoStop}%
\bibitem [{\citenamefont {Li}\ \emph {et~al.}(2019)\citenamefont {Li},
  \citenamefont {Shteynas},\ and\ \citenamefont {Ketterle}}]{floquet-heating}%
  \BibitemOpen
  \bibfield  {author} {\bibinfo {author} {\bibfnamefont {J.-R.}\ \bibnamefont
  {Li}}, \bibinfo {author} {\bibfnamefont {B.}~\bibnamefont {Shteynas}}, \ and\
  \bibinfo {author} {\bibfnamefont {W.}~\bibnamefont {Ketterle}},\ }\href
  {\doibase 10.1103/PhysRevA.100.033406} {\bibfield  {journal} {\bibinfo
  {journal} {Phys. Rev. A}\ }\textbf {\bibinfo {volume} {100}},\ \bibinfo
  {pages} {033406} (\bibinfo {year} {2019})}\BibitemShut {NoStop}%
\bibitem [{\citenamefont {Xu}\ \emph {et~al.}(2014)\citenamefont {Xu},
  \citenamefont {Qu}, \citenamefont {Gong},\ and\ \citenamefont
  {Zhang}}]{xu-mf}%
  \BibitemOpen
  \bibfield  {author} {\bibinfo {author} {\bibfnamefont {Y.}~\bibnamefont
  {Xu}}, \bibinfo {author} {\bibfnamefont {C.}~\bibnamefont {Qu}}, \bibinfo
  {author} {\bibfnamefont {M.}~\bibnamefont {Gong}}, \ and\ \bibinfo {author}
  {\bibfnamefont {C.}~\bibnamefont {Zhang}},\ }\href {\doibase
  10.1103/PhysRevA.89.013607} {\bibfield  {journal} {\bibinfo  {journal} {Phys.
  Rev. A}\ }\textbf {\bibinfo {volume} {89}},\ \bibinfo {pages} {013607}
  (\bibinfo {year} {2014})}\BibitemShut {NoStop}%
\bibitem [{\citenamefont {Zheng}\ \emph {et~al.}(2018)\citenamefont {Zheng},
  \citenamefont {Guo}, \citenamefont {Zheng},\ and\ \citenamefont
  {Zou}}]{bkt6}%
  \BibitemOpen
  \bibfield  {author} {\bibinfo {author} {\bibfnamefont {Z.-F.}\ \bibnamefont
  {Zheng}}, \bibinfo {author} {\bibfnamefont {G.-C.}\ \bibnamefont {Guo}},
  \bibinfo {author} {\bibfnamefont {Z.}~\bibnamefont {Zheng}}, \ and\ \bibinfo
  {author} {\bibfnamefont {X.-B.}\ \bibnamefont {Zou}},\ }\href {\doibase
  10.1088/1367-2630/aac7f1} {\bibfield  {journal} {\bibinfo  {journal} {New J.
  Phys.}\ }\textbf {\bibinfo {volume} {20}},\ \bibinfo {pages} {063001}
  (\bibinfo {year} {2018})}\BibitemShut {NoStop}%
\end{thebibliography}%

\end{document}